\documentclass[aps,12pt]{revtex4}
\usepackage{epsfig}
\usepackage{graphicx}
\begin{document}
\newcommand{\sn}{\rm{ sn}}
\newcommand{\cn}{\rm{ cn}}
\newcommand{\dn}{\rm{ dn}}
\newcommand{\R}{\rm{Re}}
\newcommand{\I}{\rm{Im}}
\title{Exact Periodic Solutions of Shell Models of Turbulence}
\author{Poul Olesen and Mogens H. Jensen}
\affiliation{The Niels Bohr Institute, Blegdamsvej 17, DK-2100
Copenhagen, Denmark}
\date{\today}
\begin{abstract}
We derive exact analytical solutions of the GOY shell
model of turbulence. In the absence of forcing and viscosity
we obtain closed form solutions in terms of Jacobi elliptic
functions. With three shells the model is integrable. In the case
of many shells, we derive exact recursion relations for the
amplitudes of the Jacobi functions relating the different shells
and we obtain a Kolmogorov solution in the limit of infinitely 
many shells. For the special
case of six and nine shells, these recursions relations are solved
giving specific analytic solutions. Some of these solutions
are stable whereas others are unstable. All our predictions
are substantiated by numerical simulations of the GOY shell
model. From these simulations we also identify cases where
the models exhibits transitions to chaotic states lying
on strange attractors or ergodic energy surfaces. 
\end{abstract}

\maketitle

\section{Introduction}

The behavior of turbulent systems is fairly well understood
from numerical simulations of the Navier-Stokes equations
but a deep analytical understanding of fully developed
turbulence based on these equations is still lacking
(see for instance \cite{frisch} and references therein). A fully
developed turbulent state is obtained when the Reynolds number
of the flow grows towards infinity which for instance can be
realized in the limit of vanishing viscosity. This limit of
the Navier-Stokes equations is known to be singular and
when the viscosity is identically zero the system is usually
termed the Euler equations (without external forcing) which 
are believed to exhibit a singularity in a finite time.

Direct numerical simulations of the Navier-Stokes equations at
high Reynolds numbers are generally very cumbersome and in the 
last decades
shell models of turbulence have become strong alternatives
to direct simulations \cite{Gledzer,OY,JPV,mogens}. 
Indeed, using shell model
approaches one easily simulates Reynolds numbers of the order
from $10^{10}$ up to $10^{14}$. Shells models provide excellent agreement
with statistical experimental data,
such as structure functions, probability distribution functions, etc
\cite{mogens}.
In contrast shell models generally fail to provide knowledge 
about geometrical
structures as these models are too purely resolved in Fourier
space.

In this paper we take our starting point in shell models with relatively
few shells in the Euler case of vanishing viscosity and no external
forcing. This might not
be a very realistic case for a real fluid flow but we are nevertheless
able to solve the shell model exactly in this limit. In the case of
three shells we find the full dynamical solutions in terms of Jacobi
elliptic functions. We furthermore show that the solutions are
stable as is confirmed by numerical simulations. In addition, we find
that no other solutions are possible meaning that the GOY shell model
with three shells is here shown to be fully integrable. If we consider this
system as a low-dimensional version of the Euler equations (possessing
the same basic structure and symmetries as the real Euler equations),
our results
lead to the interesting conclusion that despite the driving there
are no finite time singularities for such a system.

It is well known that the GOY shell model possesses a period 
three symmetry \cite{Gledzer,OY,JPV,mogens}.
We therefore attack the model with six and nine shells 
and find again 
analytical solutions is term of Jacobi elliptic functions. We obtain
the interesting results that some of these periodic solutions are unstable
whereas others are stable. This observation is very much related to
previous studies on periodic solutions in the GOY model \cite{GOYper},
in other turbulent systems 
\cite{KawaharaKida,Veen1,Veen2,TohItano,Waleffe,Eckhardt}
and in the Kuramoto-Shivashinsky equation \cite{KS1,KS2}.
Kawahara and Kida identified periodic orbits in plane Couette 
flow \cite{KawaharaKida} and van Veen and Kida have looked
for periodic orbits in a symmetry reduced models of fully developed
turbulence \cite{Veen1,Veen2}. Toh and Itano \cite{TohItano} and
Walaffe \cite{Waleffe} have identified periodic structures in
channel flows and Eckhardt et al has studied pipe flows \cite{Eckhardt}.
The Kuramoto-Shivashinsky system is a field equation
that does not model a turbulent fluid but nevertheless
possesses chaotic states developed from non-linear mode interactions.
Ref. \cite{KS1} identified a multitude of unstable
periodic solution in the KS equation. 
However, Frisch and co-workers found that the phase
space of the KS equations contains small isolated areas which include
stable periodic solutions \cite{KS2}. 
It is nevertheless believed that a chaotic
attractor of a large dimension in a high-dimentional space can be though
of as composed of an ordered infinity of unstable periodic orbits
(``Hopf's last hope", see \cite{CE,KS1}).
Indeed long unstable periodic orbits have been obtained for the GOY model
subjected to forcing and viscosity \cite{GOYper} thus supporting
this picture of a strange attractor. Our results therefore bridge
these observations by finding exact analytical periodic solutions to
the unforced GOY model which appear both to be stable and unstable.
It is important to note that our analytical solutions possess a phase
freedom so our results define a continuous infinity of exact solutions
for the GOY model. Since we shall show (see Appendix A) that for 
infinitely many shells a Kolmogorov solution is obtained, this means
that the periodic solutions are physically important and constitute
a non-trivial part of the phase space.

We also note that one can
obtains similar Jacobi type solutions for the Sabra 
shell model \cite{sabra1}
although the specific recursion relations for the amplitudes
of the Jacobi elliptic functions possess different conjugations
(see Appendix B).
For the Sabra model self-similar quasi-soliton solutions 
have been identified in Ref. \cite{sabra2}.

When a small forcing term is added we find
that some but not all of the stable solutions become unstable.
Numerically we observe that the dynamics either
becomes chaotic on a strange attractor in the 12-dimensional space or
quasiperiodic with a regular time evolution. The transition can be
driven by varying the initial condition in the six'th shell and we
show numerically examples of such transitions.

\section{Solution of a three-shell model}

In this section we shall find solutions to the GOY-model with three shells
for the cases with no forcing where the phases are constant, and where the 
phases are variable. This model is integrable.
Further, we also consider the model with (an imaginary) force,
and find a solution. The main point in this investigation is that
it will serve as a warm up exercise for the solution with an arbitrary
number of shells given in the next section with further details
in Appendix A.

\subsection{Constant phases}

The GOY model is
a rough approximation to the Navier-Stokes equations and is
formulated on a discrete set of $k$-values, $k_n=r^n$.
In term of a complex Fourier mode, $u_n$, of the velocity field
the model reads
\begin{equation}
(\frac{d}{ dt}+\nu k_n^2 ) \ u_n \ =
  i \,k_n (  u^*_{n+1} u^*_{n+2} \, - \, \frac{\delta}{r}
u^*_{n-1} u^*_{n+1} \, - \,
\frac{1-\delta}{r^2} \,   u^*_{n-1} u^*_{n-2})  \ + \ f \delta_{n,1}
\end{equation}
with suitable boundary conditions \cite{Gledzer,OY,JPV,mogens}.
$f$ is an external, constant forcing (here on the first mode)
and $\nu$ is the kinematic viscosity. 
{}From these equations we have energy $\sum |u_n|^2$ as 
well as helicity $\sum (-1)^n k_n |u_n|^2$ 
(for 3D where $\delta=1-1/r$) and enstrophy $\sum k_n^2|u_n|^2$
(for 2D where $\delta=1+1/r^2$) conservations \cite{kada1,kada2,bruno}.

The mathematics of the three shell case has been studied before \cite{mhd}
in a reduced version of 2D MHD with interactions between only three
wave vectors. The structure of these equations is the same as the 
three-shell GOY model discussed below. However, the physics of
the model in ref.
\cite{mhd} is in 2D and is not the same as the GOY model, 
where we can study 
different dimensions via different conservation laws.

Initially, we shall for simplicity solve the three-shell model 
without any force or viscosity (i.e. $f=\nu=0$) and
with constant phases of the velocity functions $u_n$. It is then easy to see 
that the sum of these constant phases must be $\pi/2~
{\rm mod} ~\pi$. Therefore we replace
$u_1$ by $iu_1$, and take the new $u_1$ as well as $u_2$ and $u_3$ to be real.
The three-shell equations then become
\begin{equation}
\frac{du_1}{dt}=ru_2(t)~u_3(t),
\label{1}
\end{equation}
\begin{equation}
\frac{du_2}{dt}=-r\delta~u_1(t)~u_3(t),
\label{2}
\end{equation}
and
\begin{equation}
\frac{du_3}{dt}=-r(1-\delta)~u_1(t)~u_2(t).
\label{3}
\end{equation}
{}From these equations we have energy conservation as well as helicity
(3D) or enstrophy (2D) conservations. However, 
the three-shell model has two other conservation laws. Combining (\ref{2}) and
(\ref{3}) we easily find,
\begin{equation}
u_3(t)^2=L+\frac{1-\delta}{\delta}~u_2(t)^2.
\label{4}
\end{equation}
Here $L$ is a constant which is determined by the initial conditions. In a 
similar manner we obtain from eqs. (\ref{1}) and (\ref{2})
\begin{equation}
u_1(t)^2=M-\frac{1}{\delta}~u_2(t)^2,
\label{5}
\end{equation}
where $M$ is again a constant determined by the initial conditions. Thus,
\begin{equation}
L=u_3(0)^2-\frac{1-\delta}{\delta}~u_2(0)^2,~~M=u_1(0)^2+\frac{1}{\delta}~
u_2(0)^2,
\label{initial}
\end{equation}
We note that $L+M$ is the total energy.

The conservation laws (\ref{4}) and (\ref{5}) imply that the three shell model
with constant phases and no forcing is completely integrable (see eq. 
(\ref{master}) below). Hence there is no chaotic behavior. Later we
shall show that even if a constant imaginary forcing is included, the
model is fully integrable.

{}From (\ref{2}) we have 
\begin{equation}
\frac{du_2}{u_1u_3}=-r\delta~dt.
\label{6}
\end{equation}
By means of eqs. (\ref{4}) and (\ref{5}) we can express $u_1$ and $u_3$ in 
terms of $u_2$ and eq. (\ref{6}) then gives
\begin{equation}
\int_{u_2(0)}^{u_2(t)}\frac{du_2}{\sqrt{\left(L+\frac{1-\delta}{\delta}u_2^2
\right)\left(M-\frac{1}
{\delta}u_2^2\right)}}=-r\delta t.
\label{master}
\end{equation}
This solution is thus in general an elliptic integral. If we 
substitute \footnote{For the moment we disregard the possibility $L=0$.}
$y=u_2/\sqrt{\delta M}={\rm cn}x$, where cn is the Jacobi elliptic function,
we obtain
\begin{eqnarray}
u_2(t)&=&\sqrt{\delta |u_1(0)|^2+|u_2(0)|^2}\nonumber \\
&\times& {\rm cn}\left(\sqrt{\delta}r\sqrt{|u_3(0)|^2+(1-\delta)|u_1(0)|^2}~t+
{\rm cn}^{-1}\left(\frac{u_2(0)}{\sqrt{\delta |u_1(0)|^2+|u_2(0)|^2}}\right)
\right),
\label{n1}
\end{eqnarray}
where the modulus of the Jacobi elliptic function is given by
\begin{equation}
k^2=\frac{(1-\delta)M}{L+(1-\delta)M}=(1-\delta )\frac{|u_1(0)|^2+1/\delta ~
|u_2(0)|^2}{|u_3(0)|^2+
(1-\delta )|u_1(0)|^2}.
\label{k^2}
\end{equation}
Also, cn$^{-1}$ is the inverse Jacobi function. The Jacobi functions depend on 
the elliptic integrals ($k'^2=1-k^2$)
\begin{equation}
K\equiv K(k)=\int_0^{\pi/2}\frac{d\theta}
{\sqrt{1-k^2~\sin^2\theta}},~~~K'\equiv K(k').
\label{K}
\end{equation}
  By use of the conservation laws eqs. (\ref{4}) and (\ref{5}) we
can easily find the other $u'$s,
\begin{eqnarray}
u_1(t)&=&i\sqrt{|u_1(0)|^2+1/\delta ~|u_2(0)|^2}\nonumber \\
&\times&{\rm sn}\left(\sqrt{\delta}r\sqrt{|u_3(0)|^2+(1-\delta)|u_1(0)|^2}~t+
{\rm cn}^{-1}\left(\frac{u_2(0)}{\sqrt{\delta |u_1(0)|^2+
|u_2(0)|^2}}\right)\right),
\label{n2}
\end{eqnarray}
where we reintroduced the phase in $u_1$, and
\begin{eqnarray}
u_3(t)&=&\sqrt{|u_3(0)|^2+(1-\delta )|u_1(0)|^2}\nonumber \\
&\times&{\rm dn}\left(\sqrt{\delta}r\sqrt{|u_3(0)|^2+(1-\delta)|u_1(0)|^2}~t+
{\rm cn}^{-1}\left(\frac{u_2(0)}{\sqrt{\delta |u_1(0)|^2+
|u_2(0)|^2}}\right)\right).
\label{n3}
\end{eqnarray}
Here we used that the Jacobi elliptic functions satisfy
\begin{equation}
{\rm sn}^2~ x+{\rm cn}^2~ x=1,~~{\rm dn}^2~ x+k^2{\rm sn}^2~x=1,
\label{dependence}
\end{equation}
and their derivatives satisfy
\begin{equation}
\frac{d~{\rm sn}~x}{dx}={\rm cn}~x{\rm dn}~x,~~\frac{d~{\rm cn}~x}{dx}=
-{\rm sn}~x{\rm dn}~x,~{\rm and}~~\frac{d~{\rm dn}~x}{dx}=
-k^2{\rm sn}~x{\rm dn}~x.
\label{derivative}
\end{equation}
These relations clearly shows the intimate connection between the Jacobi
elliptic functions and the three-shell model.
In the next sections
we shall see that this remark can be generalized to an arbitrary number
of shells.

As to the integrability of three-shell model, 
one can argue geometrically
as follows: Since the total energy 
is conserved, the time evolution is confined
to the surface of a sphere in $R^3$.
Now take any one of the other two conserved 
quantities eqs.(\ref{4},\ref{5}).
This corresponds to a cylinder along the $u_3$-axis with
elliptical cross section. The intersection between this
cylinder and the sphere gives then the line along which
the time evolution can proceed.
The second conserved quantity could limit everything
to points, but it does not do so, because it is not
linearly independent of the other two.
So this is then the geometrical origin of the solutions in terms
of elliptic functions.

We mention that the elliptic functions can be defined by (see \cite{WW})
\begin{eqnarray}
&&u=\int_0^{{\rm sn}(u,k)}\frac{dt}{\sqrt{(1-t^2)(1-k^2t^2)}},~u=\int_{{\rm cn}
(u,k)}^1\frac{dt}{\sqrt{(1-t^2)(k'^2+k^2t^2)}},\nonumber \\
&&{\rm and}~~u=\int_{{\rm dn}(u,k)}^1\frac{dt}{\sqrt{(1-t^2)(t^2-k'^2)}},
\label{definition}
\end{eqnarray}
They have product representations and Fourier expansions
\begin{equation}
{\rm sn}(u,k)=2q^{1/4}k^{-1/2}\sin x\prod_1^\infty~\frac{1-2q^{2n}
\cos 2x+q^{4n}}{
1-2q^{2n-1}\cos 2x+q^{4n-2}}=\frac{2\pi}{Kk}\sum_0^\infty \frac{q^{n+1/2}\sin 
(2n+1)x}{1-q^{2n+1}},
\label{sn}
\end{equation}
\begin{equation}
{\rm cn}(u,k)=2q^{1/4}k^{-1/2}k'^{1/2}\cos x\prod_1^\infty~
\frac{1+2q^{2n}\cos 2x+q^{4n}}{
1-2q^{2n-1}\cos 2x+q^{4n-2}}=\frac{2\pi}{Kk}\sum_0^\infty \frac{q^{n+1/2}\cos 
(2n+1)x}{1+q^{2n-1}},
\label{cn}
\end{equation}
and
\begin{equation}
{\rm dn}(u,k)=k'^{1/2}\prod_1^\infty~
\frac{1+2q^{2n-1}\cos 2x+q^{4n-2}}{
1-2q^{2n-1}\cos 2x+q^{4n-2}}=\frac{\pi}{2K}+\frac{2\pi}{K}\sum_1^\infty 
\frac{q^{n}\cos 2nx}{1+q^{2n}},
\label{dn}
\end{equation}
Here $x=\pi u/2K$ and $q=\exp (-\pi K'/K)$. The constants $K$ and $K'$ can be
obtained from the hypergeometric function
\begin{equation}
K=\frac{\pi}{2}~_2F_1 (\frac{1}{2},\frac{1}{2};1;k^2)~~{\rm and}~~
K'=\frac{\pi}{2}~_2F_1 (\frac{1}{2},\frac{1}{2};1;1-k^2).
\label{hyper}
\end{equation}
The hypergeometric function $_2F_1(a,b;c;z)$ is analytic in the plane cut from
from $z=1$ to $\infty$. Since $k^2$ depends on $\delta$ through 
eq. (\ref{k^2}) all results can be analytically continued in the cut
plane from 2D to 3D.
\begin{figure}[htbp]
\hbox{
{
\epsfig{width=.32\textwidth,file=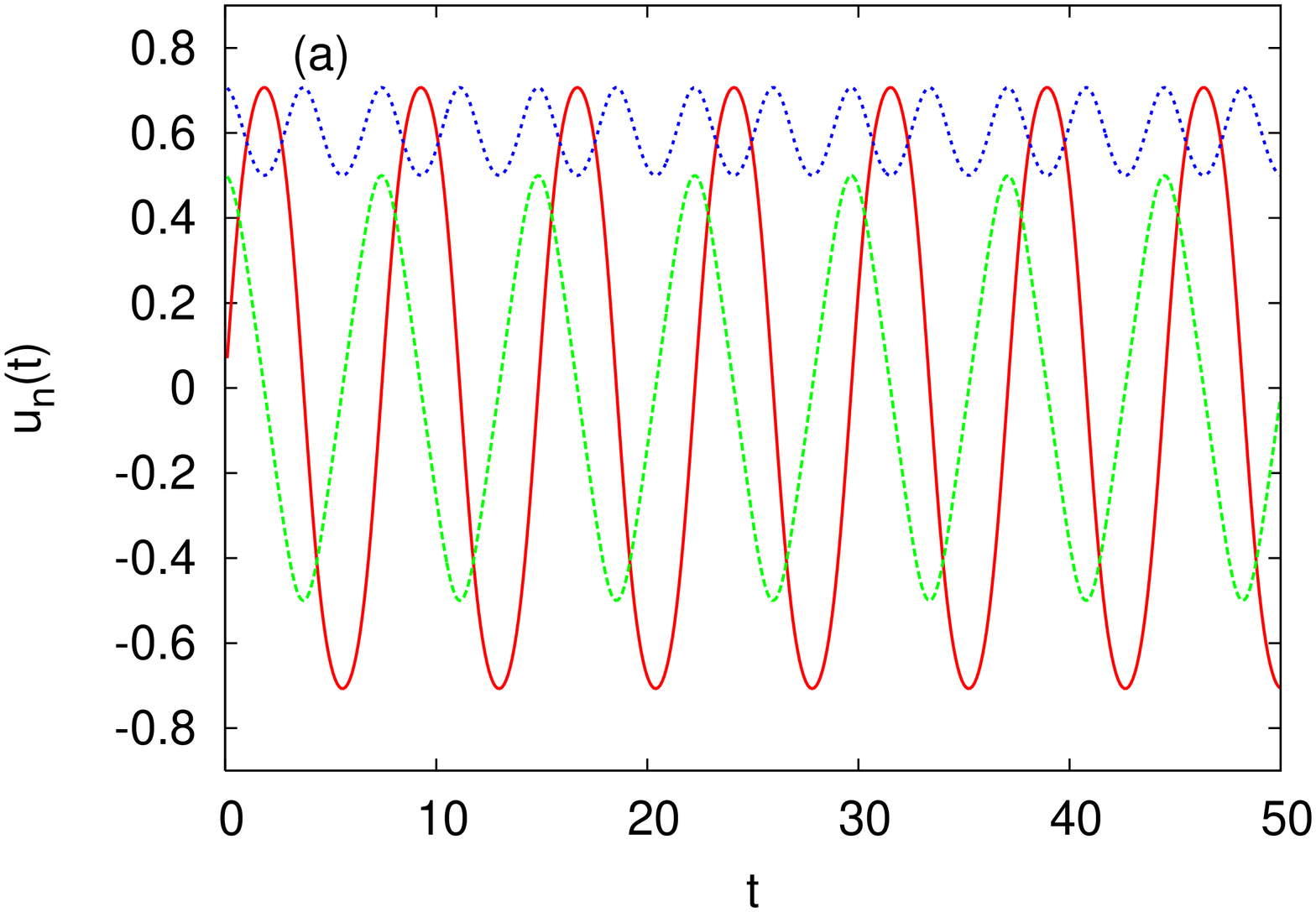,angle=-90}
}
{
\epsfig{width=.32\textwidth,file=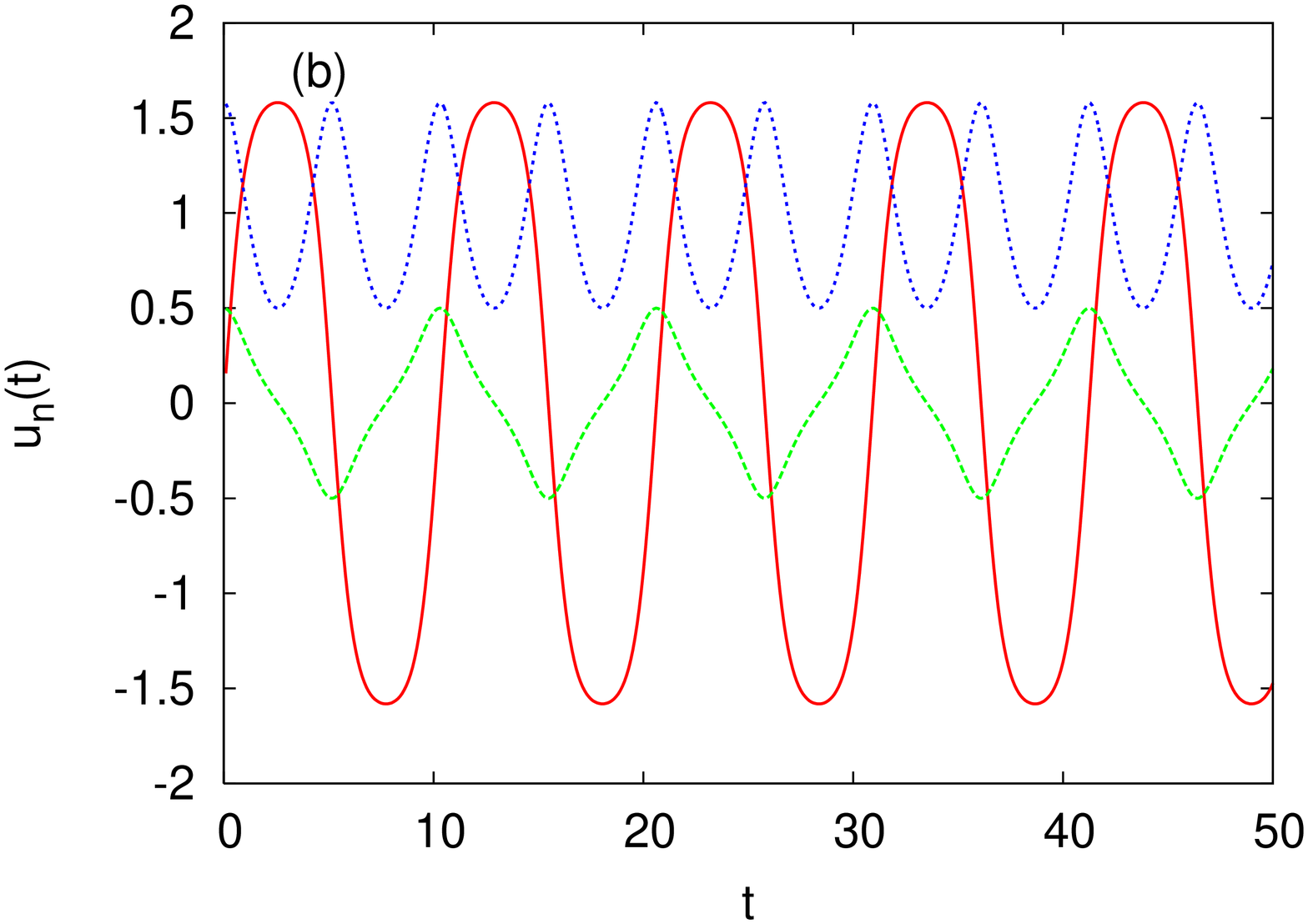,angle=-90}
}}
\caption{Plot of numerical solutions to a GOY shell
model with three shells and complex amplitudes $u_n(t)$ in the absence
of forcing and viscosity. Plotted are:
$\R~u_1(t) ({\rm red}),~ Re~u_2(t) ({\rm green}),~Im~u_3(t) ({\rm blue})$ 
with the initial conditions: 
$\R~u_1(0)=0,~Im~u_1(0)=0,~Re~u_2(0)=0.5,~Im~u_2(0)=0,
~Re~u_3(0)=0,~Im~u_3(0)=0.5/\sqrt{\delta}$. 
a): $\delta = 0.5$, b): $\delta = 0.1$. The curves correspond to the exact 
solutions Eq. (\ref{simplesolution}).}
     \label{fig:1}
\end{figure}

It should be noted that for small $k$ the Fourier expansions in practice only
contain very few terms, since $q$ is small. This also applied when $k$ is
not so small. For example, if $k^2 = \frac{1}{2}$, then $K = K'$
and hence $q=e^{-\pi}$, so $q^n$ is fast decreasing. 
However, when $k$ approaches 1, a 
large number of terms 
are needed in the Fourier transforms because in
this case $q$ is not small.
This follows because $q$ can be expressed in terms of $k$ by the formula
\begin{equation}
q=\lambda +2 \lambda^5+15 \lambda^9+150\lambda^{13}+...,~~\lambda=
\frac{1}{2}~\frac{1-(1-k^2)^{1/4}}{1+(1-k^2)^{1/4}}.
\end{equation}

In the case $L=0$, i.e. $u_3(0)^2=(1-\delta )/\delta~u_2(0)^2$, 
the solution (\ref{master}) does not give an oscillating solution any longer.
Instead one obtains
\begin{eqnarray}
&&u_2(t)=u_2(0)\nonumber \\
&&\times\frac{2~\sqrt{u_1(0)^2+u_2(0)^2/\delta}~e^{-\sqrt{M\delta (1-\delta)}
~rt}}{\sqrt{u_1(0)^2+u_2(0)^2/\delta}+
u_1(0)+(\sqrt{u_1(0)^2+u_2(0)^2/\delta}-u_1(0))~e^{-2\sqrt{M\delta (1-\delta)}
~rt}}.
\label{exp}
\end{eqnarray}
In this case $k=1$, and obviously an infinite number of Fourier modes
are needed to produce the exponentially decreasing behavior (\ref{exp}).
{}From eq. (\ref{exp}) one can easily obtain $u_1$ by use of the conservation
law and (\ref{5}),
\begin{eqnarray}
&&u_1(t)=\sqrt{u_1(0)^2+u_2(0)^2/\delta}\nonumber \\
&&\frac{\sqrt{u_1(0)^2+u_2(0)^2/\delta}+u_1(0)-(\sqrt{u_1(0)^2+u_2(0)^2/\delta}
-u_1(0))~e^{-2\sqrt{M\delta (1-\delta)}~rt}}{\sqrt{u_1(0)^2+u_2(0)^2/\delta}+
u_1(0)+(\sqrt{u_1(0)^2+u_2(0)^2/\delta}-
u_1(0))~e^{-2\sqrt{M\delta (1-\delta)}~rt}}.
\end{eqnarray}
Thus, at $t\rightarrow\infty$ all the energy has gone to the $u_1-$mode.

In three dimensions $1-\delta$ is positive, whereas in two dimension this
quantity is negative. From (\ref{k^2}) it then follows that $k^2 > 0$ 
in 3D,
whereas $k^2<0$ in 2D if $|u_3(0)|^2>(\delta-1)|u_1(0)|^2$. The Jacobi elliptic
functions can be continued to negative $k^2$, so this does not pose any 
problem. The continued functions are given by
\begin{eqnarray}
{\rm sn}(x,ik)&=&\frac{1}{\sqrt{1+k^2}}~\frac{{\rm sn}(x\sqrt{1+k^2},\frac{k}
{\sqrt{1+k^2}})}{{\rm dn}(x\sqrt{1+k^2},\frac{k}
{\sqrt{1+k^2}})},~~{\rm cn}(x,ik)=\frac{{\rm cn}(x\sqrt{1+k^2},\frac{k}
{\sqrt{1+k^2}})}{{\rm dn}(x\sqrt{1+k^2},\frac{k}{\sqrt{1+k^2}})},\nonumber \\
{\rm dn}(x,ik)&=&\frac{1}{{\rm dn}(x\sqrt{1+k^2},\frac{k}{\sqrt{1+k^2}})}.
\end{eqnarray}
For $k\rightarrow 0\pm$ these functions are continuous.

The solutions (\ref{n1})-(\ref{n3}) simplify if $u_1(0)=0$ and $u_2(0)^2
/\delta=u_3(0)^2$. From (\ref{k^2}) we see that in this case
$k^2=1-\delta$, and
\begin{equation}
u_1(t)=i\frac{a}{\sqrt{\delta}}~{\rm sn} (art,\sqrt{1-\delta}),~~u_2(t)=
a~{\rm cn} (art,\sqrt{1-\delta}), 
~~u_3(t)=\frac{a}{\sqrt{\delta}}~{\rm dn} (art,\sqrt{1-\delta}).
\label{simplesolution}
\end{equation}
Here $a=u_2(0)$.
In Fig. \ref{fig:1} we show these exact solutions by direct
numerical integration of the GOY model with three shells
(with specific initial conditions). Note
the familiar behavior of the Jacobi elliptic functions. Two
different values of the parameter $\delta$ are shown in
order to vary the elliptic functions parameter $k^2 =1-\delta$. 
The solutions are stable and the numerical integrations
can proceed indefinitely. 

We also mention that if $k=0$, corresponding to $\delta=1$, sn and cn become
sin and cos, respectively, and dn approaches the constant one. From eq. 
(\ref{k^2}) we see that $k^2=0$ if $\delta =1$. This case has been studied
(including viscosity) in \cite{po}.

\subsection{Solution with variable phases}

We shall now discuss the case of variable phases. In
forming the energy or other conserved quantities they do not enter
{\it directly}. However, in this section we shall see that the phases do play
a dynamical role. Letting the phases be allowed to vary a new solution emerges,
which can still be expressed in terms of Jacobian elliptic function, but
which has one more free parameter than the solutions (\ref{n1})-(\ref{n3}).

The three-shell model  with 
variable phases defined by $u_n=|u_n|e^{i\phi_n}$ can be rewritten as 
six equations,
\begin{eqnarray}
&&\frac{d|u_1|}{rdt}=|u_2||u_3|~\sin (\phi_1+\phi_2+\phi_3),~\frac{d|u_2|}
{rdt}=-\delta |u_1||u_3|~\sin (\phi_1+\phi_2+\phi_3),\nonumber \\
&&{\rm and}~\frac{d|u_3|}{rdt}=
-(1-\delta )|u_2||u_1|~\sin (\phi_1+\phi_2+\phi_3),
\label{modulus}
\end{eqnarray}
as well as
\begin{eqnarray}
&&|u_1|~\frac{d\phi_1}{rdt}=|u_2||u_3|~\cos (\phi_1+\phi_2+\phi_3),~
|u_2|~\frac{d\phi_2}{rdt}=-\delta |u_1||u_3|~
\cos (\phi_1+\phi_2+\phi_3),\nonumber \\
&&{\rm and}~|u_3|~\frac{d\phi_3}{rdt}=-(1-\delta)|u_1||u_2|~\cos (\phi_1+
\phi_2+\phi_3).
\label{phase}
\end{eqnarray}
{}From the last equations we easily derive
\begin{equation}
\frac{d(\phi_1+\phi_2+\phi_3)}{rdt}=-\cos (\phi_1+\phi_2+\phi_3)~\left[-\frac
{|u_2||u_3|}{|u_1|}+\delta\frac{|u_1||u_3|}{|u_2|}+(1-\delta )\frac{|u_1||u_2|}
{|u_3|}\right],
\label{100}
\end{equation}
On the left we can use eqs. (\ref{modulus}) in the square bracket to get
\begin{equation}
\frac{d(\phi_1+\phi_2+\phi_3)}{dt}=\cot (\phi_1+\phi_2+\phi_3)~\frac{d\ln 
(|u_1||u_2||u_3|)}{dt},
\label{200}
\end{equation}
from which we get
\begin{equation}
\cos (\phi_1+\phi_2+\phi_3)=\frac{N}{|u_1||u_2||u_3|},
\label{300}
\end{equation}
where $N$ is a constant. We see that if the phases are constants, i.e. $N=0$, 
their sum must be $\pi/2$ mod $\pi$ (i.e. $\pi/2,~3\pi/2$,...), as already 
used in the previous subsection.

{}From the cosine in (\ref{300}) we can compute the sine and insert it
in eqs. (\ref{modulus}). By this procedure we arrive at
\begin{eqnarray}
&&\frac{d|u_1|}{rdt}=\frac{1}{|u_1|}~\sqrt{|u_1|^2|u_2|^2|u_3|^2-N^2},~~
\frac{d|u_2|}{rdt}=-\delta~\frac{1}{|u_2|}~\sqrt{|u_1|^2|u_2|^2|u_3|^2-
N^2}\nonumber \\
&&{\rm and}~\frac{d|u_3|}{rdt}=-(1-\delta )~\frac{1}{|u_3|}~\sqrt{|u_1|^2
|u_2|^2|u_3|^2-N^2}. 
\label{600}
\end{eqnarray}
In this equation we have inserted a positive square root, corresponding
to a positive sine. Since the left hand sides of the eqs. (\ref{600}) can
also be negative in some range of $t$ values, it follows that for such
values of $t$ a negative sign should be chosen for in front of the
square roots in eqs. (\ref{600}), 
corresponding to negative values of the sine.
In order not to make the notation too clumsy
we do not indicate this explicitly.
Further from eq. (\ref{modulus}) we can easily derive 
like eqs. (\ref{4}), (\ref{5})
\begin{equation}
|u_3|^2=L+\frac{1-\delta}{\delta}|u_2|^2 ~~{\rm and}~~|u_1|^2=M-
\frac{1}{\delta}|u_2|^2.
\label{700}
\end{equation}
This allows us to express any $|u_n|$ in terms of the other two $|u|'$s.
Using this in eqs. (\ref{600}) we obtain for example for $|u_2|$ the equation
of motion,
\begin{equation}
\frac{d|u_2|^2}{2\delta\sqrt{|u_2|^2 (M-|u_2|^2/\delta )(L+|u_2|^2~
(1-\delta)/\delta)-N^2}}=-d(rt).
\label{800}
\end{equation}
This is again an elliptic type of integral. However, it is slightly more
complicated than the solutions considered in the previous subsection.

Inside the square root we have a polynomial of third degree in $|u_2|^2$.
If we introduce the zeros of this polynomial,
this is the prototype of an integral which can be inverted by
the use of Theta functions \cite{WW}. However, 
instead of seeking the most general solution,
in the following we shall study only a limited range of parameters 
corresponding to a limited range of the initial values of the $u'$s.

To obtain a relatively simple solution we must remember that the 
Jacobi elliptic 
functions are not really independent, as is seen from eq. (\ref{dependence}).
Therefore one could attempt to make an ansatz in such a manner that
inside the square root we have three squares of elliptic functions,
which requires
\begin{equation}
|u_1|^2=A_1+B_1~{\rm sn}^2(art),~|u_2|^2=A_2+B_2~{\rm cn}^2(art),~{\rm and}~
|u_3|^2=A_3+B_3~{\rm dn}^2(art).
\label{ansatz1}
\end{equation}
With a suitable choice of the $A$ and $B$ constants the square root
in eq. (\ref{800})
could then produce the product ${\rm sn}(at)~{\rm cn}(at)~{\rm dn}(at)$
to be matched by the derivative in the numerator,
$d{\rm cn}^2 (at)=-2{\rm cn}(at)~{\rm sn}(at)~{\rm dn}(at)~dt$ 
(see eq. (\ref{derivative})).

Instead of proceeding from the integral (\ref{800}) it is actually easier to 
insert the ansatz (\ref{ansatz1}) directly in the equations of motion
(\ref{600}). This gives a number of algebraic equations. When we differentiate
the functions (\ref{ansatz1}) in order to insert them on the left
hand sides in (\ref{600}) we get a result proportional to $B_n~ {\rm sn}(art)~
{\rm cn}(art)~{\rm dn}(art)/|u_n|$ with $n=1,2,3$ by use of eqs. (\ref
{derivative}). To match the right hand sides of eqs. (\ref{600})
we therefore need
\begin{equation}
\sqrt{|u_1|^2|u_2|^2|u_3|^2-N^2}\propto~ {\rm sn}(art){\rm cn}(art){\rm dn}
(art).
\end{equation}
This means that the following conditions must be satisfied \footnote{In 
arriving at these results we have first considered the term in the products of
the functions (\ref{ansatz1}) which obviously produces the wanted
result. Then we have combined the other terms to be coefficients of sn$^4(art)$, 
sn$^2(art)$ and sn$^0(art)$ by use of the relations (\ref{dependence}). These
coefficients must vanish, which is the requirements in (\ref{condition1})-
(\ref{condition3}).},
\begin{equation}
aB_1=\sqrt{B_1B_2B_3},~~aB_2=\delta~\sqrt{B_1B_2B_3}, ~~k^2aB_3=(1-\delta )
\sqrt{B_1B_2B_3}.
\label{B}
\end{equation}
As before in eq. (\ref{simplesolution}), 
we take $k^2=1-\delta$. Then the restrictions (\ref{B})
are easily solved
\begin{equation}
B_1=a^2/\delta,~B_2=a^2,~~B_3=a^2/\delta.
\label{Bcond}
\end{equation}
The remaining conditions then become
\begin{equation}
0=-A_3a^4/\delta-A_2(1-\delta )a^4/\delta^2+A_1(1-\delta )a^4/\delta,
\label{condition1}
\end{equation}
which is the requirement that the coefficient of sn$^4(art)$ vanish, and
\begin{equation}
N^2=A_1A_2A_3+A_1A_3a^2+A_1A_2a^2/\delta+A_1a^4/\delta,
\label{condition2}
\end{equation}
which expresses the condition that the coefficient of sn$^0(art)$ vanishes,
and
\begin{equation}
0=A_2A_3a^2/\delta-A_1A_3a^2+A_3a^4/\delta-A_1A_2(1-\delta )a^2/\delta
+A_2a^4/\delta^2-A_1a^4/\delta-A_1(1-\delta )a^4/\delta,
\label{condition3}
\end{equation}
which expresses the condition that the coefficient of sn$^2(art)$ vanishes.

The solutions of (\ref{condition1})-(\ref{condition3}) should be such that 
the square root on the right hand sides of eq. (\ref{600}) should be real.
This is not a trivial requirement. One can study this question in general by 
first solving for $A_3$ in terms of $A_1,A_2$ by use of eq.
(\ref{condition1}), and (\ref{condition3}) then becomes a second order equation
which gives $A_2$ in terms of $A_1$. In this way we obtain
\begin{equation}
A_3=(1-\delta)(A_1-A_2/\delta),
\label{emil}
\end{equation}
\begin{equation}
A_2=\frac{\delta}{2(1-\delta)}\left(a^2+(1-\delta)A_1\pm\sqrt{a^4+(1-\delta)A_1
\left(2a^2\frac{\delta-2}{\delta}-3(1-\delta)A_1\right)}~\right),
\label{emil2}
\end{equation}
as well as
\begin{equation}
N^2=A_1^3~\delta (1-\delta)+A_1^2a^2 (2-\delta)+A_1a^4/\delta.
\label{nn}
\end{equation}
The last restriction can be compared to eq. (\ref{300}), which expresses
the constant $N$ in terms of the initial values of the phases and
the $|u|'$s. Using eq. (\ref{Bcond}) to find the initial $|u|'$s and
comparing this to eq. (\ref{nn}) we see that the ansatz (\ref{ansatz1})
is valid only when cosine of the sum of the initial phases is
equal to plus/minus one, i.e.
\begin{equation}
\phi_1(0)+\phi_2(0)+\phi_3(0)=0~({\rm mod}~\pi).
\end{equation}
This does not mean that the phases at later times are restricted in this way.

The requirements (\ref{emil}),(\ref{emil2}) and (\ref{nn}) do
not give acceptable (real) results for all values of $A_1$ and $a$,
since we need to ensure that in (\ref{nn}) $N^2>0$ and that the square
root in (\ref{emil2}) should be real.
Let us therefore give a numerical example where the results are good. We
have found that if $\delta =1-1/r=1/2$ (3D, with the often used value $r=2$)
and $a=1$ 
the following values are completely consistent (the square root is real!),
\begin{equation}
A_1=0.309,~~A_2=0.577,~~A_3=-0.423,~~N^2=0.7685.
\end{equation}
For initial values of the phases and the $|u|'$s which are not covered by the 
ansatz (\ref{ansatz1}) one must go back to the standard elliptic integral 
(\ref{800}), and treat it in other ways than done here.

The solution (\ref{ansatz1}) can be used to obtain an integral representation
for the phases. We can insert the cosine from eq. (\ref{300}) in the equations
of motion for the phases (\ref{phase}), and we get
\begin{equation}
\phi_1 (t)=\phi_1(0)+N~\int^t_0\frac{dt}{A_1+(a^2/\delta)~{\rm sn}^2(art)},
\end{equation} 
with similar expressions for the other phases.

It should also be mentioned that it is possible to take into account
the viscosity terms in the equations for the phases. The result is
\begin{equation}
\frac{d\phi_1}{dt}=\frac{N}{|u_1|^2}~e^{-\nu (k_1^2+k_2^2+k_3^2)t},
\end{equation}
with similar expressions for the other phases. Unfortunately we have
not been able to solve for the modulus $|u_1|$ when viscosity is included.

In conclusion we have seen that if the phases are allowed to vary,
this situation gives rize to the new solutions (\ref{ansatz1}). So,
although the phases do not appear in the energy and other conserved quantities
and one might therefore consider them unphysical, this is not true
because of their dynamical importance. As the phases are allowed 
to vary continuously this indeed outlines an infinity of exact periodic
solutions to the GOY models with three shells.

\subsection{Inclusion of a constant force}

We shall now consider the case where a constant force ($f$ is taken to be 
imaginary, $f\rightarrow if$)
is included in the first shell, so that (\ref{1}) reads
\begin{equation}
\frac{du_1}{dt}=ru_2(t)~u_3(t)+f,
\label{glemt}
\end{equation}
whereas the other two equations (\ref{2}) and (\ref{3}) are unchanged.
For this reason we derive like in eq. (\ref{4})
\begin{equation}
u_3(t)^2=L+\frac{1-\delta}{\delta}~u_2(t)^2.
\label{f1}
\end{equation}
However, eq. (\ref{5}) is now replaced by
\begin{equation}
\frac{1}{2}~u_1(t)^2-f\int^t_0 dt'~u_1(t')=\frac{1}{2}~M-\frac{1}{2\delta}~
u_2(t)^2.
\label{f2}
\end{equation}
Here the initial conditions again determine the constants $M$ and $L$. 
The integral
on the left hand side can be determined from eq. (\ref{2}) and (\ref{f1}),
\begin{equation}
-r\delta ~u_1=\frac{1}{u_3}\frac{du_2}{dt}=\frac{du_2/dt}{\sqrt{L+
\frac{1-\delta}{\delta}~u_2^2}}.
\label{f3}
\end{equation}
Thus,
\begin{equation}
-r\delta~\int_0^tdt'~u_1(t')=\sqrt{\frac{\delta}{1-\delta}}~\ln \frac{u_2(t)
+\sqrt{u_2(t)^2+\frac{L\delta}{1-\delta}}}{u_2(0)+\sqrt{u_2(0)^2+
\frac{L\delta}{1-\delta}}}.
\label{f4}
\end{equation}
Inserting this in eq. (\ref{2}) we obtain by use of eq. (\ref{f1})
\begin{equation}
-r\delta~ t=\int\frac{du_2}{u_1u_3}=\int_{u_2(0)}^{u_2(t)}
\frac{du_2}{\sqrt{\left(L+\frac{1-\delta}
{\delta}~u_2^2\right)\left(M-\frac{1}{\delta}u_2^2-f\frac{1}
{r\sqrt{\delta (1-\delta)}}~\ln \frac{u_2
+\sqrt{u_2^2+\frac{L\delta}{1-\delta}}}{u_2(0)+\sqrt{u_2(0)^2+
\frac{L\delta}{1-\delta}}}~\right)}}.
\label{f5}
\end{equation}
This in principle expresses $u_2$ as a function of $t$, and $u_1$ can
then be obtained from (\ref{f2}), whereas $u_3$ can be obtained from
(\ref{f1}). However, due to the logarithm in the denominator 
we cannot use elliptic functions to invert eq. (\ref{f5}).

If $L=0$ and $f/r$ is large and positive, it is 
possible to obtain an asymptotic 
expression for the integral (\ref{f5}) by introducing the variable
\begin{equation}
z=M-\frac{1}{\delta}u_2^2-\frac{2f}{r\sqrt{\delta (1-\delta)}}~\ln
\frac{u_2}{u_2(0)}.
\label{z} 
\end{equation}
By a partial integration one then obtains
\begin{eqnarray}
&&-\sqrt{\delta(1-\delta)}~rt=-\int_0^{z_0}\frac{dz}{\sqrt{z}
\left(2u_2^2/\delta+2f/(r\sqrt{\delta(1-\delta)}\right)}\nonumber\\
&&=-\left[\frac{2\sqrt{z}}{2u_2^2/\delta +2f/(r\sqrt{\delta (1-\delta)})}
\right]_0^{z_0}
~\left(1+O\left(\frac{1}{f}\right)\right).
\end{eqnarray}
Here $z_0$ is obtained from the substitution (\ref{z}) just by inserting the 
upper limit $u_2(t)$ instead of the integration variable $u_2$.
With the assumption $u_2\ll f |\ln u_2|$ 
we can invert this expression and obtain
\begin{equation}
u_2(t)\approx u_2(0)~e^{-\sqrt{\delta (1-\delta})~frt^2/2}.
\end{equation}
This expression works quite well  for large $f'$s, as one can see
by solving eq. (\ref{glemt}) numerically. 
Thus the assumption $u_2\ll f |\ln u_2|$ is self-consistent.

\begin{figure}[htbp]
\hbox{
{
\epsfig{width=.36\textwidth,file=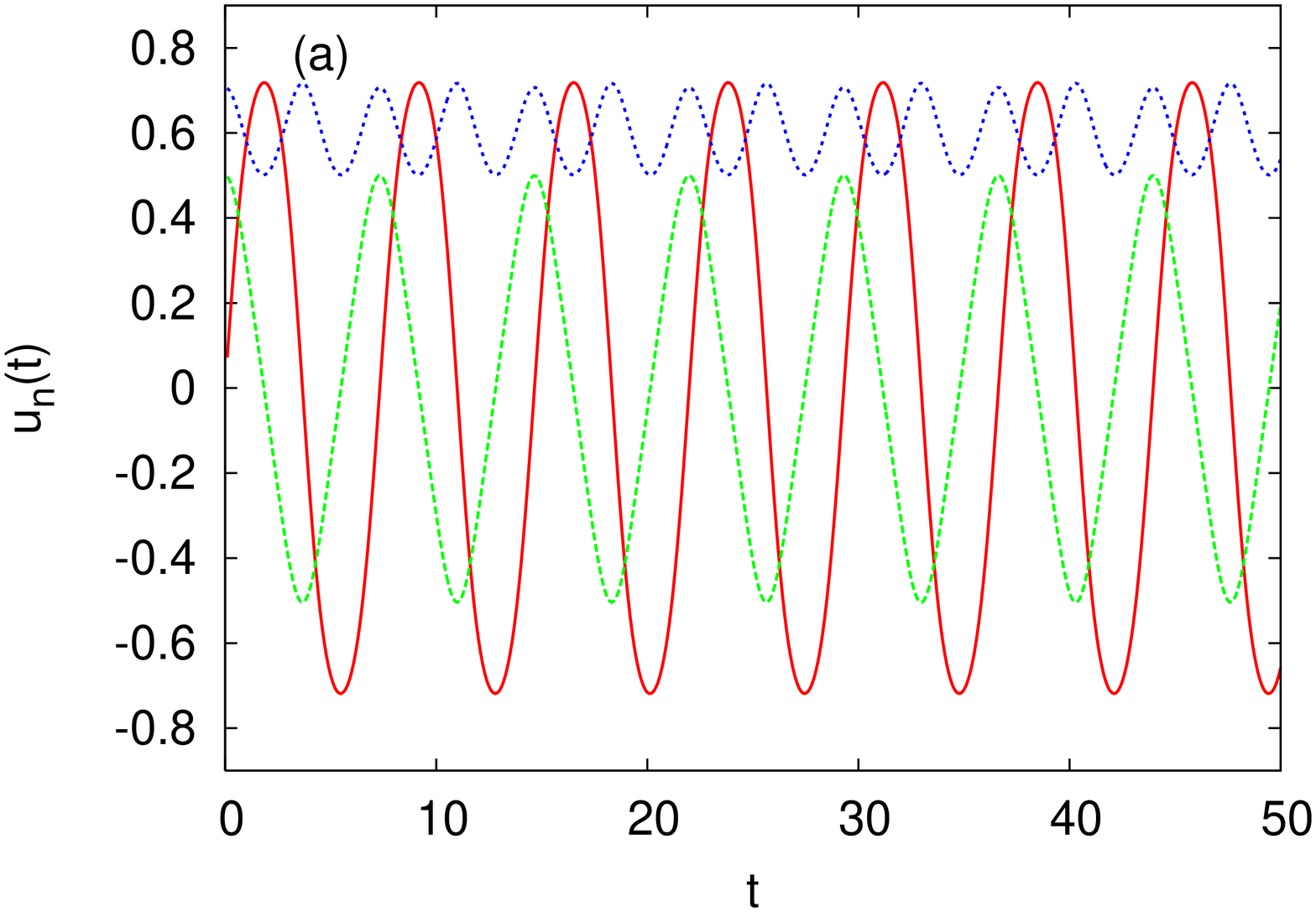,angle=-90}
}
{
\epsfig{width=.36\textwidth,file=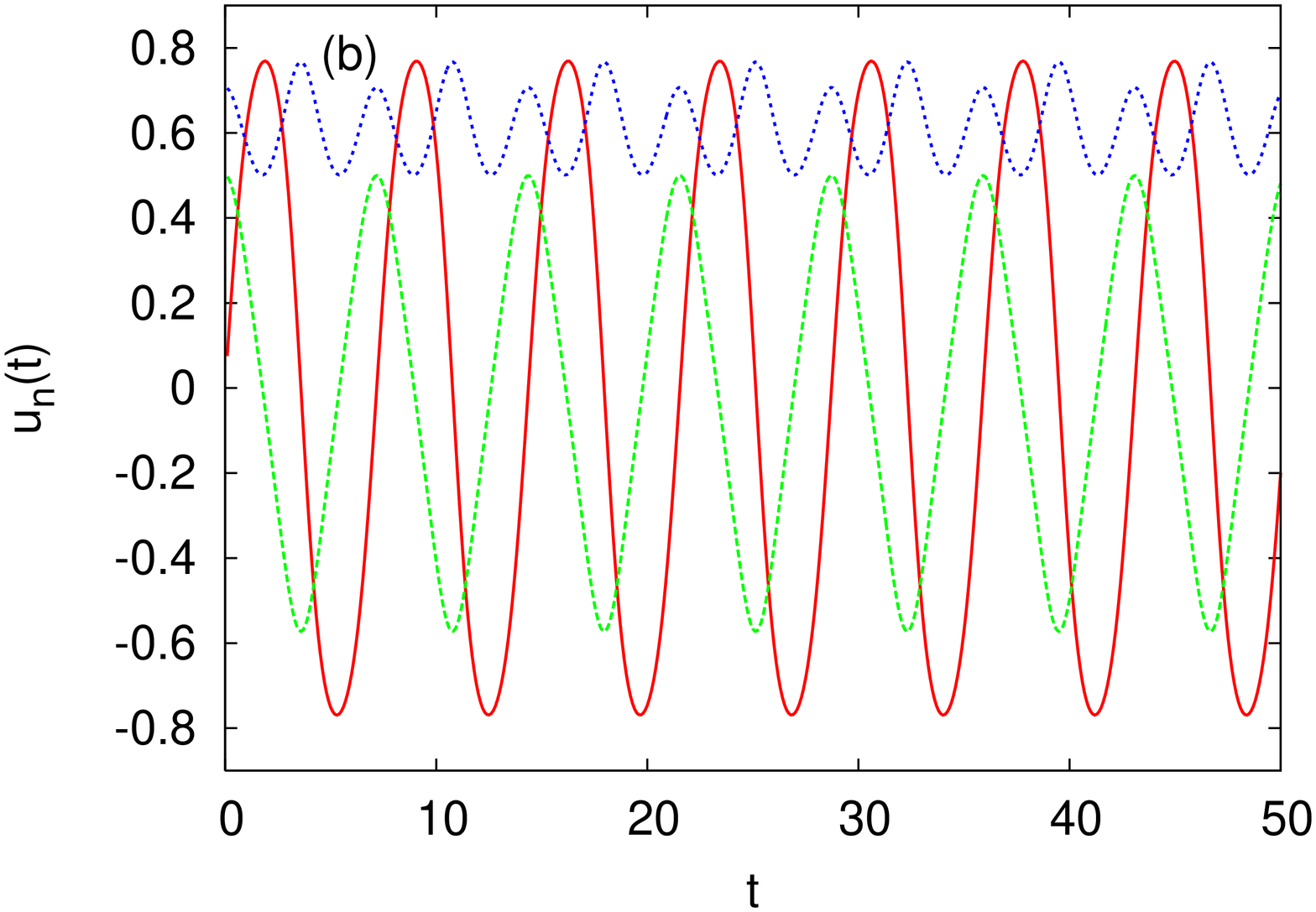,angle=-90}
}}
\caption{Plot of the Jacobi elliptic functions for three shells
with addition of a force, a): $f=(1+i)\cdot 10^{-2}$
b):  $f=(5+5i)\cdot 10^{-2}$ with otherwise the same values of
the parameters and the same initial conditions as in Fig. 1.
Note that compared to a) the
variation in the dn function has undergone a period doubling.
In both cases a),b) the solutions are stable.}
\label{fig:2}
\end{figure}

For $f<0$ we have not been able to find asymptotic results. Numerically one can
easily see that in this case $u_2(t)$ oscillates. In the Fig.
\ref{fig:2} we show numerical solutions of the GOY model with three
shells with a small forcing added. The solutions are surprisingly
still stable even though energy is now added into the system. We
observe the interesting phenomenon that the dn-function undergoes
a period doubling when the value of the forcing is changed. We
were however not able to identify a series of period doubling
eventually leading into a chaotic state.

\section{Many shells}

In this section we shall extend one of the results obtained in the last
section. We start by defining the functions
\begin{eqnarray}
Ja_1&=&{\rm sn} (art+b),~Ja_2={\rm cn}(art+b),~Ja_3={\rm dn}(art+b),\nonumber\\
Ja_4&=&{\rm sn}(art+b),~Ja_5={\rm cn}(art+b),~Ja_6={\rm dn}(art+b),\nonumber \\
Ja_7&=&{\rm sn} (art+b),...,
\end{eqnarray}
where the symbol $Ja$ stands for Jacobi. In general $Ja_{n}=Ja_{n-3}$. The 
$Ja'$s satisfy the differential equations
\begin{equation}
dJa_n/dt=\epsilon_n~Ja_{n+1}Ja_{n+2}.
\end{equation}
Here
\begin{equation}
\epsilon_n=+1 ~{\rm if}~n=1~~({\rm mod}~3),~~\epsilon_n=-1 ~{\rm if}~n=2~~(
{\rm mod}~3),~~\epsilon_n=-k^2 ~{\rm if}~n=3~~({\rm mod}~3).
\end{equation}

\begin{figure}[htbp]
\hbox{
{
\epsfig{width=.32\textwidth,file=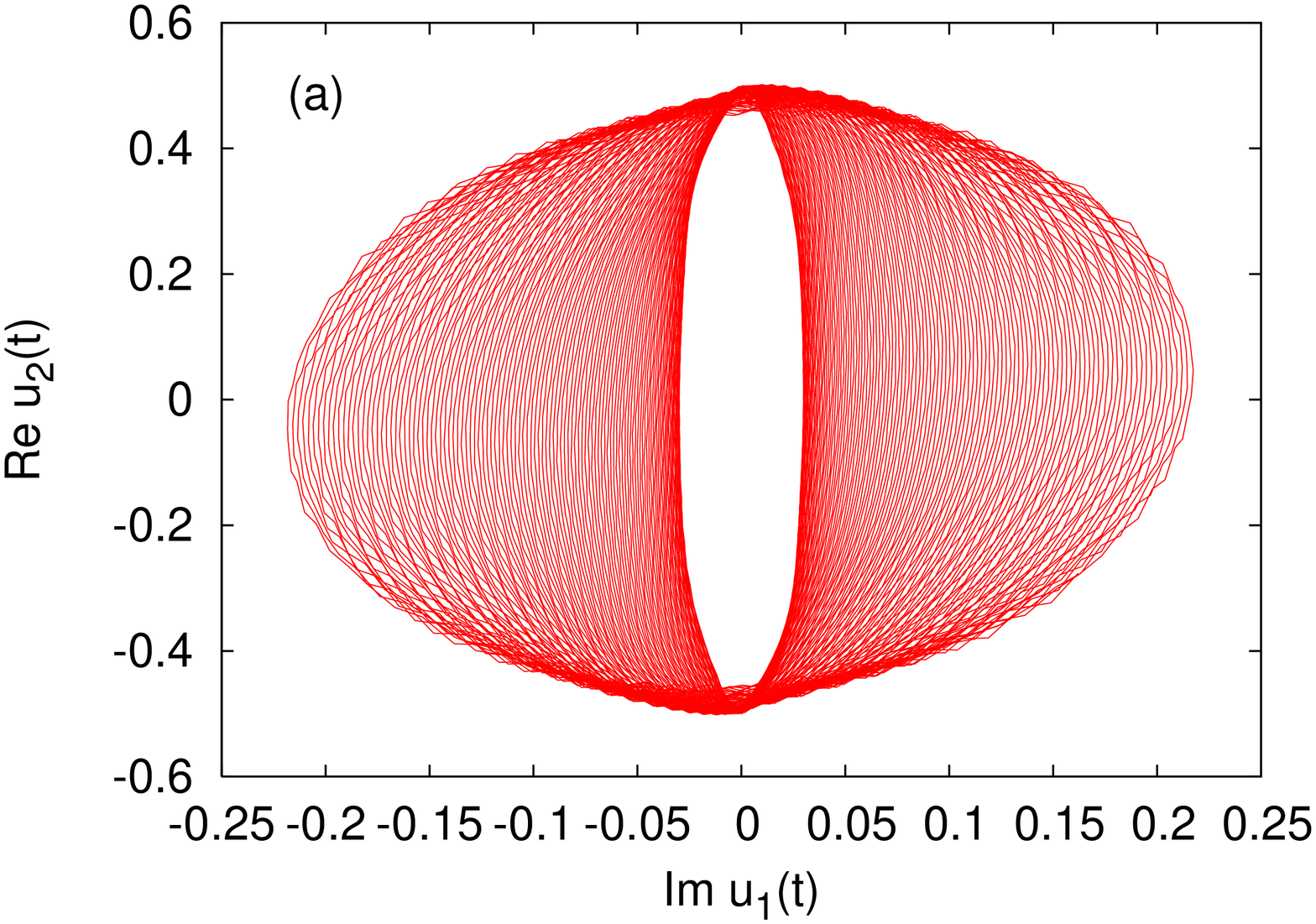,angle=-90}
}
{
\epsfig{width=.32\textwidth,file=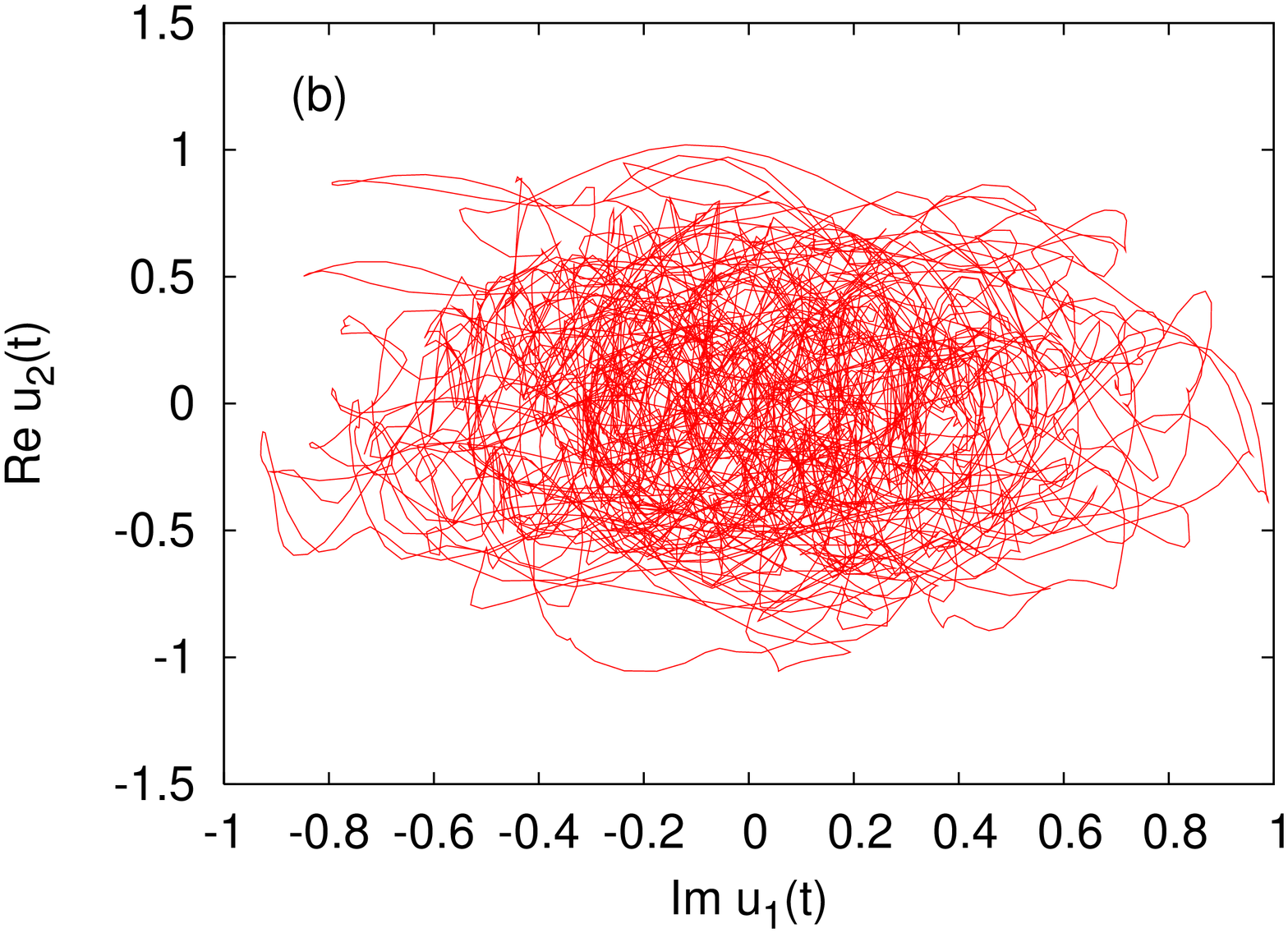,angle=-90}
}}
\caption{Numerical solutions of the unforced GOY model
six shells. The initial conditions are: 
a): $\R~ u_1(0)=~ 1.2,~ Im~ u_1(0) = 0.6,~ Re~ u_2 = 0.5,~ Im~ u_2(0) = 0.9,
~Re~ u_3 = 0.5, ~Im~ u_3(0) = 2.0, ~Re~ u_4(0) = 0.0, ~Im~ u_4 (0) = 0.707,
~Re~ u_5 = 0.0, ~Im~ u_5 = .5, ~Re~ u_6 =0.0, ~Im~ u_6 =8.0$. The solution
is given by a stable quasiperiodic trajectory.
b): 
$\R~ u_1(0)= 1.2, ~Im~ u_1(0) = 0.6, ~Re~ u_2 = 0.5, ~Im~ u_2(0) = 0.9,
~Re~ u_3 = 0.5, ~Im~ u_3(0) = 2.0, ~Re~ u_4(0) = 0.0, ~Im~ u_4 (0) = 0.707,
~Re~ u_5 = 0.0, ~Im~ u_5 = .5, ~Re~ u_6 =0.0, ~Im~ u_6 =0.5$. By just changing
the initial condition for $\I~ u_6$, the solution changes from a quasiperiodic
into a chaotic trajectory.
}
\label{fig:3}
\end{figure}

The modulus $k^2$ is to be determined by the requirement that the functions
\begin{equation}
u_n=A_n~Ja_n (art+b)
\label{ansatz}
\end{equation}
should satisfy the shell equations up to some order. Similarly, the constants 
$a, ~b$ and $A_n$ should be  restricted from this requirement. Eq. 
(\ref{ansatz}) is of course a generalization of (\ref{n1}), (\ref{n2}), and 
(\ref{n3}). As an example we show in Fig. \ref{fig:3} two solution
to the unforced GOY model with six shells. By just changing the initial
conditions in the six'th shell, the solution change from being
quasiperiodic to a chaotic signal. 
Thus in the Hamiltonian 12-dimensional phase space 
stable periodic and quasiperiodic islands are situated between
chaotic areas as is a well known scenario for Hamiltonian systems
of low dimension.

The shell model without viscosity and forcing
\begin{equation}
\frac{du_n^\star}{dt}=-ik_n\left(u_{n+1}u_{n+2}-
\frac{\delta}{r}~u_{n-1}u_{n+1}-\frac{1-\delta }{r^2}~u_{n-1}u_{n-2}\right).
\end{equation}
is satisfied exactly by the ansatz (\ref{ansatz}), because 
by direct insertion we get
\begin{eqnarray}
&&{\rm LHS}=A^\star_n a r \epsilon_n Ja_{n+1}(at)Ja_{n+2}(at)= \\
&&-ik_n\left(A_{n+1}A_{n+2}Ja_{n+1}Ja_{n+2}-\frac{\delta}{r}~A_{n-1}
A_{n+1}Ja_{n-1}Ja_{n+1}-\frac{1-\delta}{r^2}~A_{n-1}A_{n-2}Ja_{n-1}Ja_{n-2}\right)\nonumber,
\label{oo}
\end{eqnarray}
and using that the $Ja'$s are defined mod 3 in the index we get
\begin{equation}
Ja_{n-1}=Ja_{n+2} ~~{\rm(in ~second~and~third~terms)},~~Ja_{n-2}=Ja_{n+1} ~~
{\rm (in~third~term)}.
\end{equation}
which shows that the $Ja'$s cancel out on both sides of eq. (63).
The $A'$s should therefore satisfy
\begin{equation}
raA^\star_n\epsilon_n=-ik_n\left(A_{n+1}A_{n+2}-\frac{\delta}{r}A_{n-1}A_{n+1}
-\frac{1-\delta}{r^2}A_{n-1}A_{n-2}\right).
\end{equation}
This can be rewritten as a recursion relation
\begin{equation}
A_{n+2}=\frac{iraA_n^\star\epsilon_n}{k_nA_{n+1}}+\frac{\delta}{r}~A_{n-1}+
\frac{1-\delta}{r^2}~\frac{A_{n-1}A_{n-2}}{A_{n+1}}.
\label{recursion}
\end{equation}

We have not succeeded in showing that this relation always has a solution.
However, it is instructive to compare to the results in the last section.
Let us start to solve the recursion relation (\ref{recursion}),
\begin{equation}
A_3=\frac{iaA_1^\star}{A_2},
\label{aaa3}
\end{equation}
\begin{equation}
A_4=\frac{1}{rA_1^\star}~\left(\delta |A_1|^2-|A_2|^2\right),
\label{a4}
\end{equation}
and
\begin{equation}
A_5=\frac{1}{rA_2^\star(\delta |A_1|^2-|A_2|^2)}~\left(-k^2a^2|A_1|^2
+(\delta^2-\delta+1)~|A_1|^2|A_2|^2-\delta~|A_2|^4\right).
\label{a5}
\end{equation}
Subsequent coefficients get more complex. The next one is
\begin{equation}
A_6=\frac{ia(A_4^\star +r(1-\delta)A_1^\star)}{r^3A_5}+\frac{ia\delta 
A_1^\star}{r A_2},
\label{a6}
\end{equation}
Here eqs. (\ref{a4}) and (\ref{a5}) should be inserted.

To gain some insight in the recursion scheme let us start by assuming that 
there are only three shells. Then we should 
require that $A_4=0$, i.e. from (\ref{a4})
\begin{equation}
|A_2|= \sqrt{\delta}~|A_1|.
\label{a1a2}
\end{equation}
Further, from the requirement $A_5 A_4=0$ one obtains
\begin{equation}
0=-k^2a^2|A_1|^2+(\delta^2-\delta+1)~|A_1|^2|A_2|^2-\delta~|A_2|^4.
\end{equation}
Inserting (\ref{a1a2}) this gives
\begin{equation}
k^2=(1-\delta )\frac{|A_2|^2}{a^2},
\end{equation}
which is exactly the same as the previously obtained result (\ref{k^2}).

We mention that usually the GOY model is just truncated at some
shell number, in this case $n=3$. However, in dealing with the
analytic solutions this is absolutely necessary since we need 
the information contained when the two subsequent
shell amplitudes (in this case $n=$4 and 5) are put equal to zero,
leading to a value for
$k^2$, which shows that only the first amplitude is a 
free parameter. Therefore
it is necessary to proceed in the somewhat unconventional way
followed above, also when we consider an arbitrary but finite number of
shells, as we shall see in the following.
\begin{figure}[htbp]
\vbox{
{
\epsfig{width=.30\textwidth,file=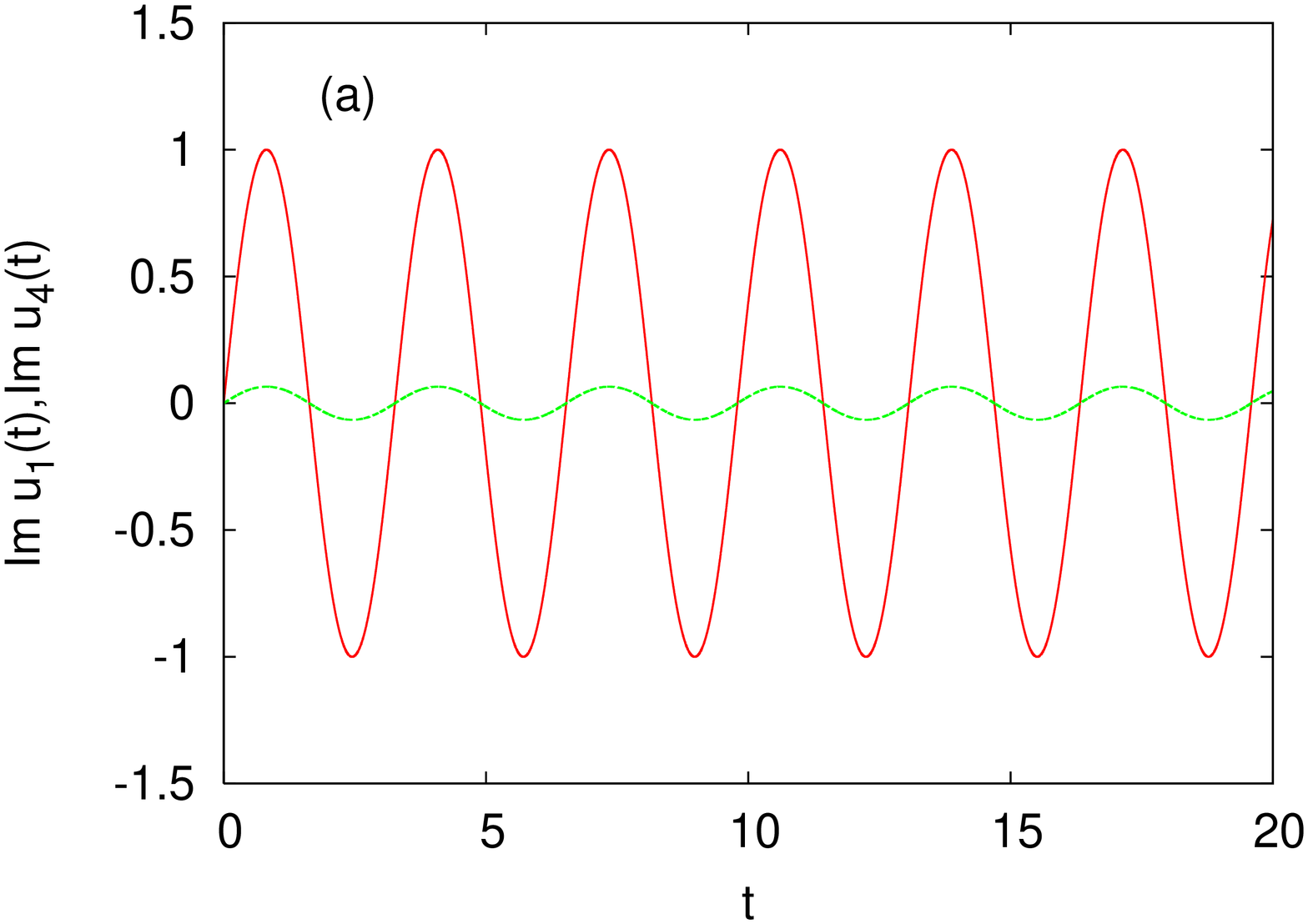,angle=-90}
}
{
\epsfig{width=.30\textwidth,file=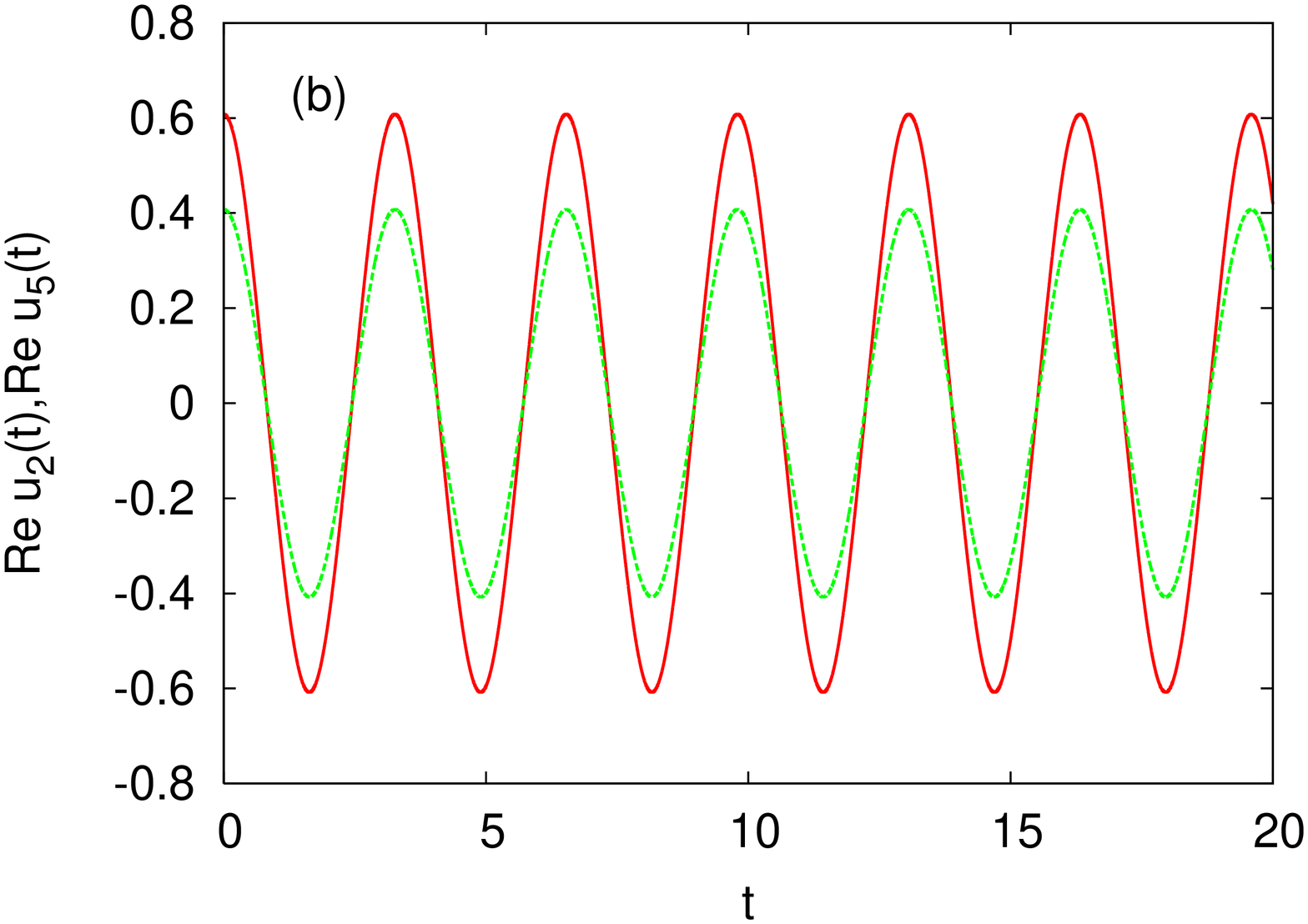,angle=-90}
}
{
\epsfig{width=.30\textwidth,file=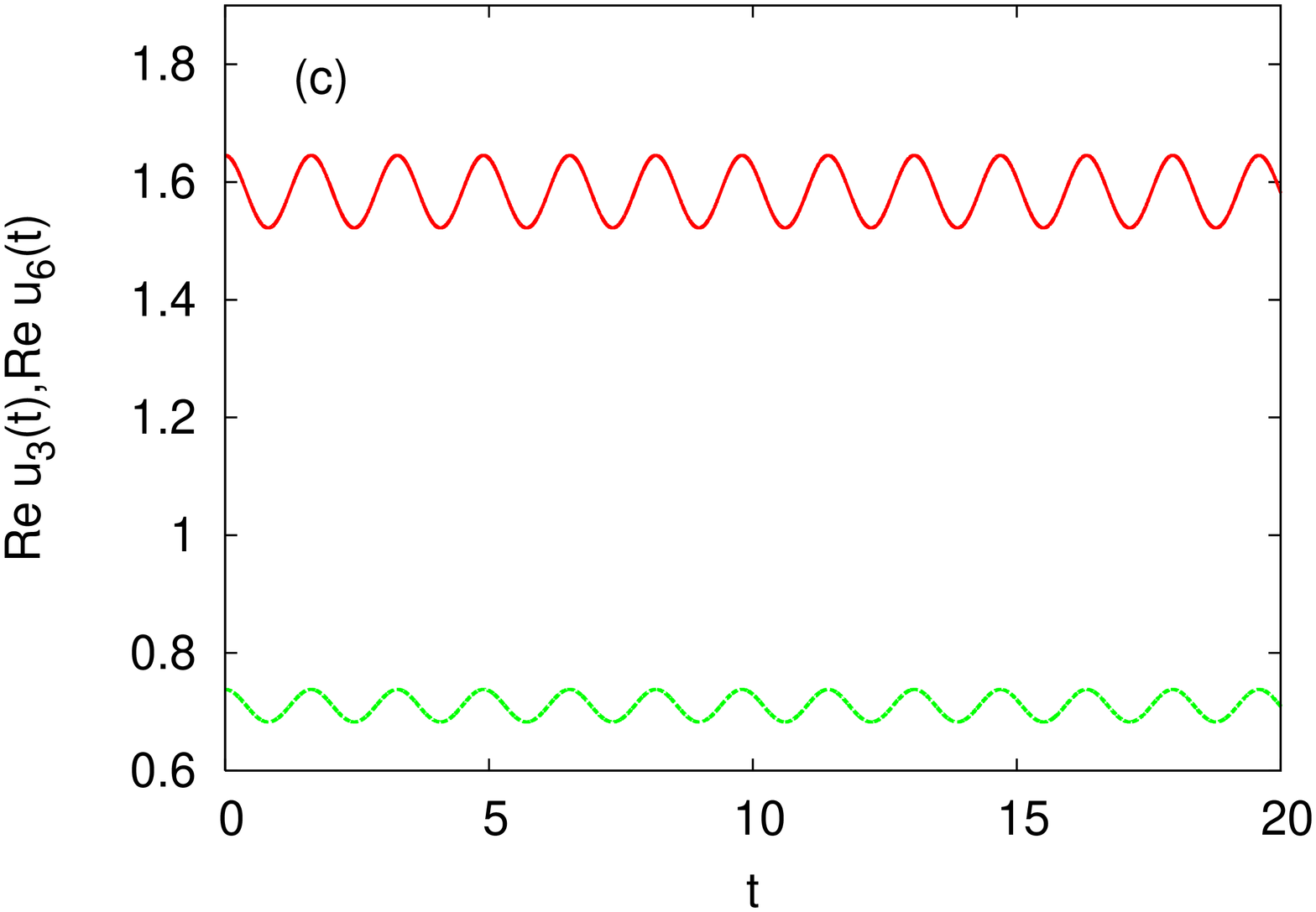,angle=-90}
}}
\caption{The solutions to the GOY model with six shells without
forcing and viscosity for $\delta =0.5$ with the following
scaling coefficients: $\alpha_4=0.06527718351453,
~~~\alpha_5=0.6704413705072, ~~~\alpha_6=0.44861490325181$
and parameter values $a=1,b=0,A_1 = 1$. This leads to the following 
initial values:
$\R~ u_1(0)= ~Im~ u_1(0) = ~Im~ u_2(0) = ~Im~ u_3(0) = ~Re~ u_4(0) = ~Im~ u_4 (0) =
~Im~ u_5(0) = ~Im~ u_6(0) =0.0, ~Re~ u_2(0) = 0.607820697, ~Re~ u_3(0) = 1.645222028,
~Re~ u_5(0) = 0.407507916,~Re~ u_6(0) =0.73807128$. 
The following shell model variables are plotted:
a): $\I~ u_1(t),~\I~ u_4(t)$; b): $\R~ u_2(t),~\R~ u_5(t)$; c):
$\R~ u_3(t),~\R~ u_6(t)$.}
\label{fig:4}
\end{figure}

With six shells \footnote{Because of the special role played by modulus 3 in 
the GOY model it is safer to end $n$ at 3,6,9,12,... ~.} we need 
\begin{equation}
A_7=0=\frac{-iaA_5^\star}{r^4A_6}+\frac{\delta}{r}A_4+\frac{1-\delta}{r^2}
\frac{iaA_1^\star A_4}{A_2A_6},
\label{e}
\end{equation}
and
\begin{equation}
A_7A_8=0=-\frac
{iak^2 A_6^\star}{r^5}+\frac{(1-\delta)}{r^2}~A_5A_4.
\label{ee}
\end{equation}
Eqs. (\ref{e}) and (\ref{ee}) can in principle be used to determine $k^2$. The 
algebra is, however, quite involved.

It is possible to simplify the rather tedious algebra by introducing
the following ansatz,
\begin{equation}
A_4=\alpha_4~A_1,~~ A_5=\alpha_5 ~A_2,~{\rm and}~~A_6=\alpha_6~A_3.
\label{ansatz5}
\end{equation}
Here we have assigned $A_1$ and $A_4$ the phase $\pi/2$, whereas the other
$A'$s are real. In the following, we take all the $A_i'$s to
be real. From eq. (\ref{a4}) we see that
\begin{equation}
A_2^2=(\delta-r\alpha_4)~A_1^2.
\label{c1}
\end{equation}
{}From eq. (\ref{a5}) we similarly get
\begin{equation}
k^2=\left(-r^2\alpha_4~\alpha_5+1-\delta +r\delta \alpha_4\right)~\frac{A_2^2}
{a^2}.
\label{c2}
\end{equation}
Next we consider the recursion relation (\ref{a6}) for $A_6$ which gives
\begin{equation}
\alpha_6=\frac{\alpha_4+r(1-\delta)}{r^3\alpha_5}+\frac{\delta}{r}.
\label{c3}
\end{equation}
The conditions for the shell model to stop at six shells are represented by
eqs. (\ref{e}) and (\ref{ee}), which lead to
\begin{equation}
\frac{\delta}{r}~\alpha_4~\alpha_6=\frac{\alpha_5}{r^4}~\left(
\delta-r\alpha_4\right)-\frac{1-\delta}{r^2}~\alpha_4,
\label{c4}
\end{equation}
and
\begin{equation}
(1-\delta)~\alpha_4~\alpha_5=\frac{\alpha_6}{r^3}~\left(-r^2\alpha_4\alpha_5+1-
\delta +r\delta~\alpha_4\right),
\label{c5}
\end{equation}
respectively. In the last equation we used $k^2$ from eq. (\ref{c2}).

{}From these three equations we can obtain the three
$\alpha'$s in terms of the
parameters $r,\delta$. This elimination procedure leads to a
sixth order equation for $\alpha_4$, and from the solutions of this
equation the other $\alpha'$s can then be obtained.
\begin{figure}[htbp]
\vbox{
{
\epsfig{width=.30\textwidth,file=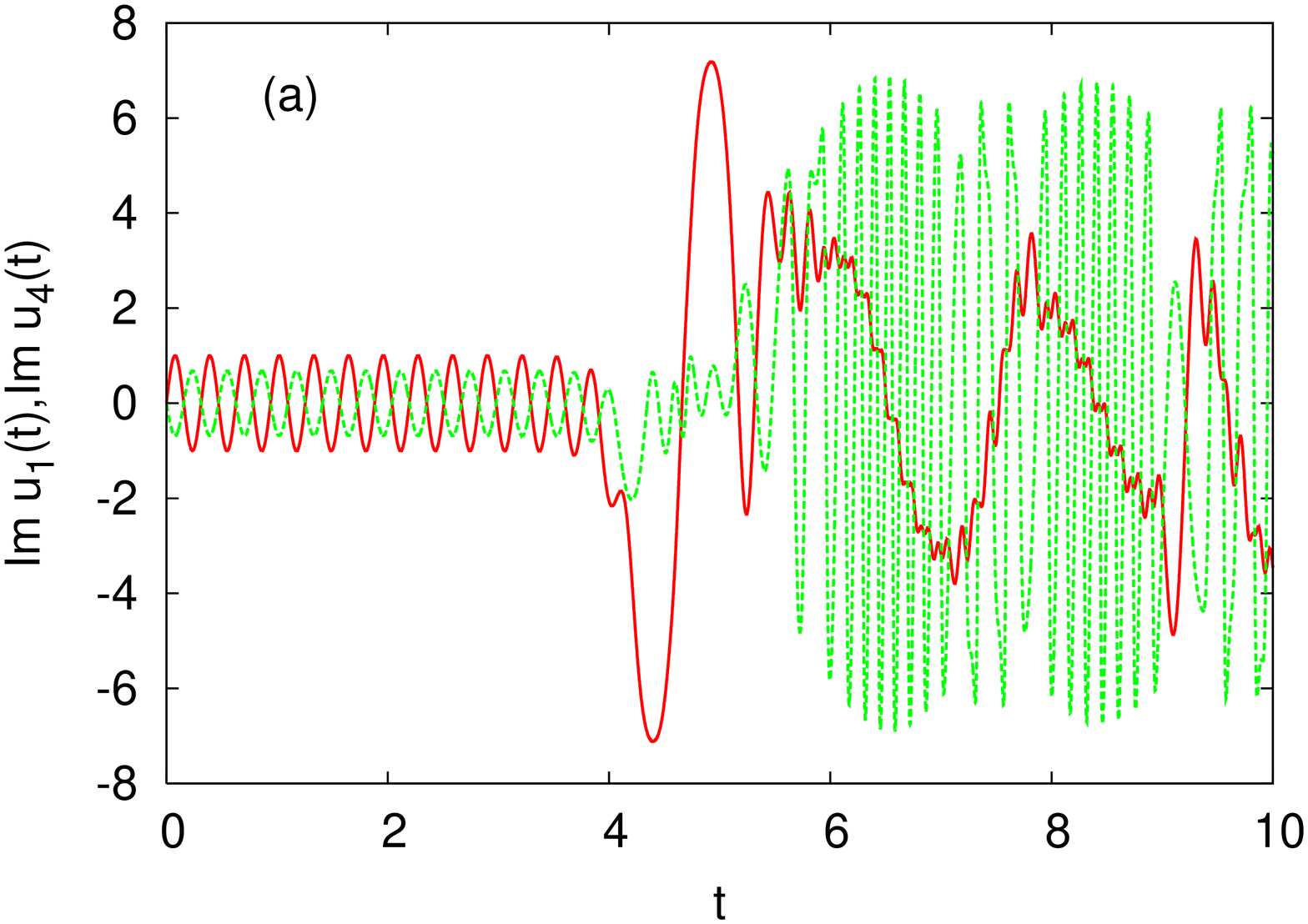,angle=-90}
}
{
\epsfig{width=.30\textwidth,file=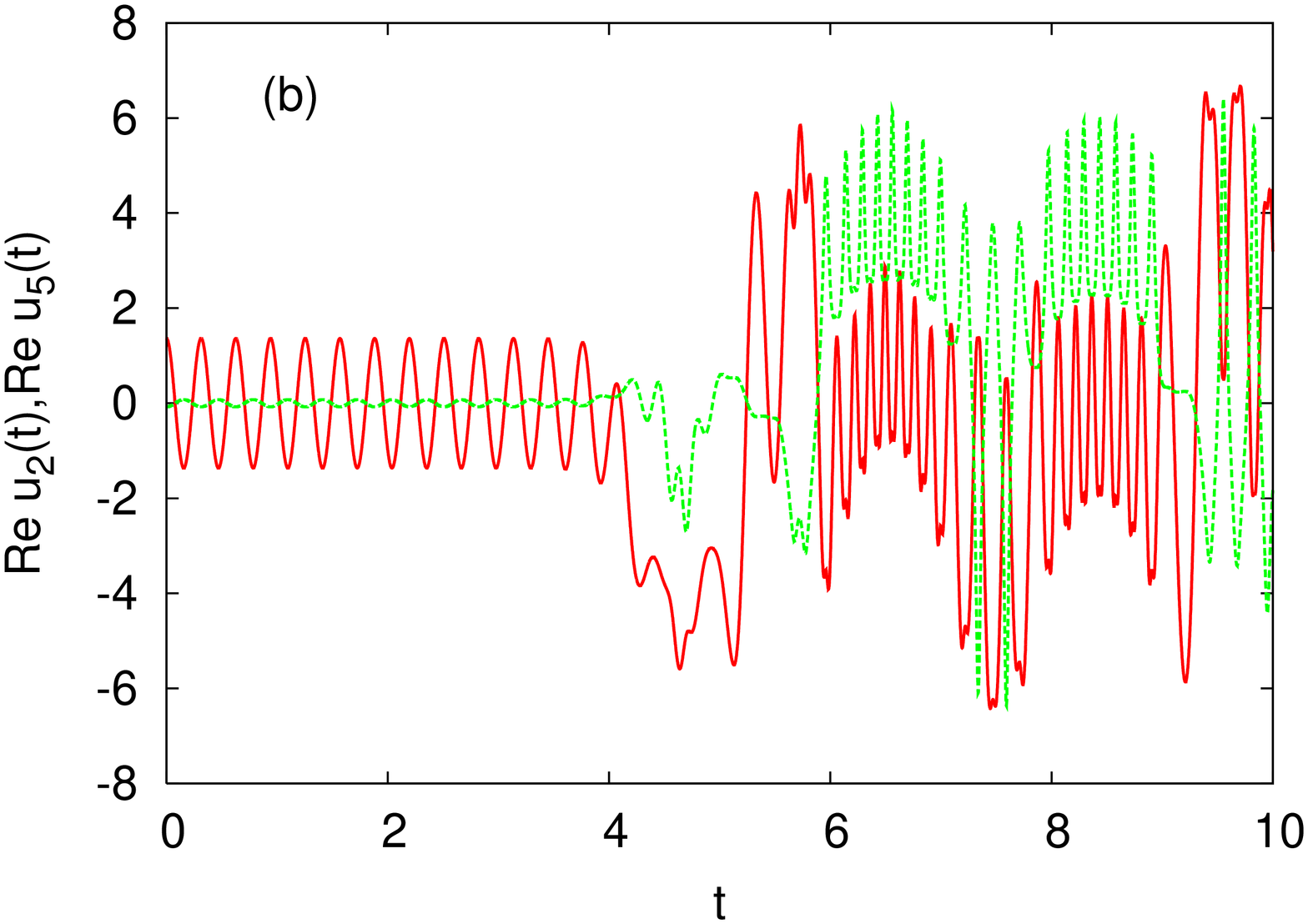,angle=-90}
}
{
\epsfig{width=.30\textwidth,file=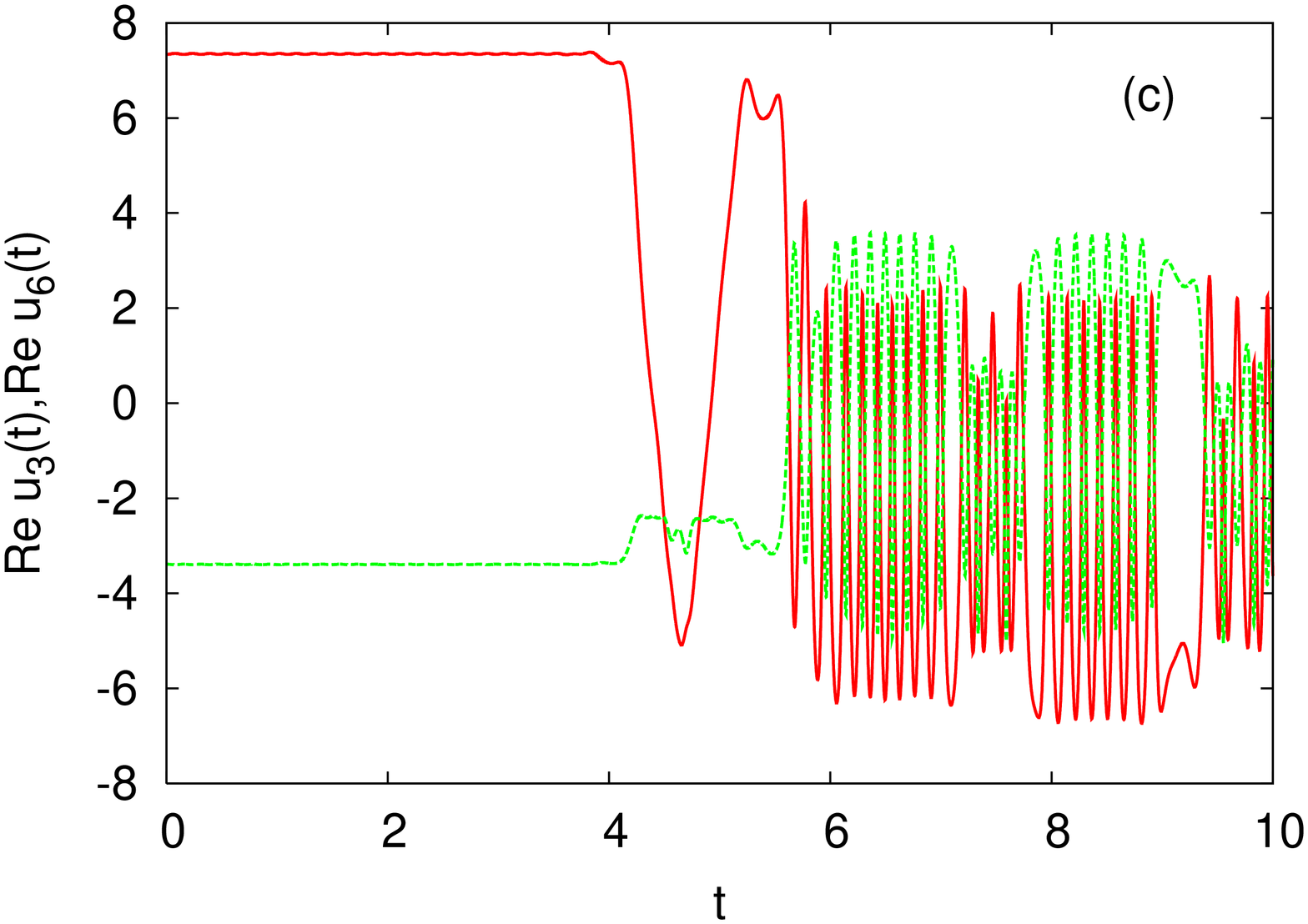,angle=-90}
}}
\caption{The solutions to the GOY model with six shells without
forcing and viscosity for $\delta =0.5$. We apply the following
scaling coefficients: $\alpha_4=-0.6788645691146,
~~~\alpha_5=-0.05642693495529,
~~~\alpha_6=-0.46139658555751$. Setting $a=10,b=0,A_1 = 1$ give 
the following initial conditions:
$\R ~u_1(0)= ~Im ~u_1(0) = ~Im ~u_2(0) = ~Im ~u_3(0) = ~Re ~u_4(0) = ~Im~ u_4 (0) =
Im ~u_5 = ~Im ~u_6 =0.0, ~Re ~u_2 = 1.362985377, ~Re ~u_3 = 7.33683587,
~Re ~u_5 = -0.076909087, ~Re ~u_6 =-3.385191019$
}
\label{fig:5}
\end{figure}

This procedure leads to extremely messy expressions which are
not useful, since we do not have analytic solutions of sixth order equations.
Instead, it is much simpler to assign numerical values for the parameters.
Taking $r=2$ and $\delta=1/2$ (3D) we obtain from eqs. (\ref{c3}), (\ref{c4})
and (\ref{c5}) by use of NSolve in Mathematica (applied to these
three simultaneous equations) the following real solutions:
\begin{eqnarray}
&&\alpha_4=0.06527718351453,~~~\alpha_5=0.6704413705072,
~~~\alpha_6=0.44861490325181\nonumber \\
&&\alpha_4=-0.6788645691146,~~~\alpha_5=-0.05642693495529,
~~~\alpha_6=-0.46139658555751,\nonumber \\
&&\alpha_4=-17.345971520666,~~~\alpha_5=1.3995954432919,
~~~\alpha_6=-1.2098836041347,\nonumber \\
&&{\rm Not~ acceptable~(see~ remark~below~eq.~} (\ref{sas})): \nonumber \\
&&\alpha_4=0.625261,~~~\alpha_5=-0.309106,~~~\alpha_6=-0.407242
\label{table}
\end{eqnarray}
{}From eq. (\ref{c1}) there are relations between $A_1$ and $A_2$, and
using the numbers in (\ref{table}) we get
\begin{equation}
A_2^2/A_1^2=\delta-r\alpha_4=0.369446,~~1.85773,~~35.1919,
\label{sas}
\end{equation}
respectively.
Here the fourth line in eq. (\ref{table}) has been ignored, since it does
not lead to a real ratio $A_2/A_1$, as assumed in the above derivation.
Thus there only three acceptable solutions.

For $k^2$ we obtain
\begin{equation}
\frac{k^2a^2}{A_2^2}=80.2634,~~~-0.33209,~~~0.390219,
\end{equation}
respectively.

In Figs. \ref{fig:4},\ref{fig:5} we show two of the exact solutions to
the model GOY with six shells as given in (\ref{table}) by
the first two set of $\alpha_n$-values. In the first case, shown
in Fig. \ref{fig:4}, we obtain the familiar Jacobi elliptic
function with the exact relationships between the amplitude
as predicted by the theory. Notice that the amplitude of 
Im $u_1$ come to exactly 1 even though the initial condition
of the mode is set to zero. These solutions are stable. In
Fig. \ref{fig:5} we show the the second analytic solution
in (\ref{table}). Note we here have set the frequency to
$a=10$ in order to obtain a value of the elliptic parameter
$k$ less than one. In this case, again, the numerical
integrations starts out tracing out the exact analytical
solutions in terms of Jacobi elliptic functions. However,
as is clear from Fig. \ref{fig:5} these solutions are
unstable since after some time they turn into chaotic trajectories.
We have experimented numerically with the accuracy of
the $\alpha_n$ coefficients and the accuracy of the integration.
As is expected, for unstable solutions in a chaotic ``sea",
the exact solution can be sustained numerically longer
the higher the accuracy.

The GOY model has been modified by changing the order of 
the complex conjugates of the velocity amplitudes resulting in the 
Sabra model \cite{sabra1}.
We have checked that the Jacobi elliptic functions are also exact
solutions to this model. One can derive recursion relations
like eqs. (\ref{c3}),(\ref{c4}),(\ref{c5}) which have a similar 
structure but the conjugations and a sign appear differently.
This is discussed in some details in Appendix B.

\begin{figure}[htbp]
\hbox{
{
\epsfig{width=.32\textwidth,file=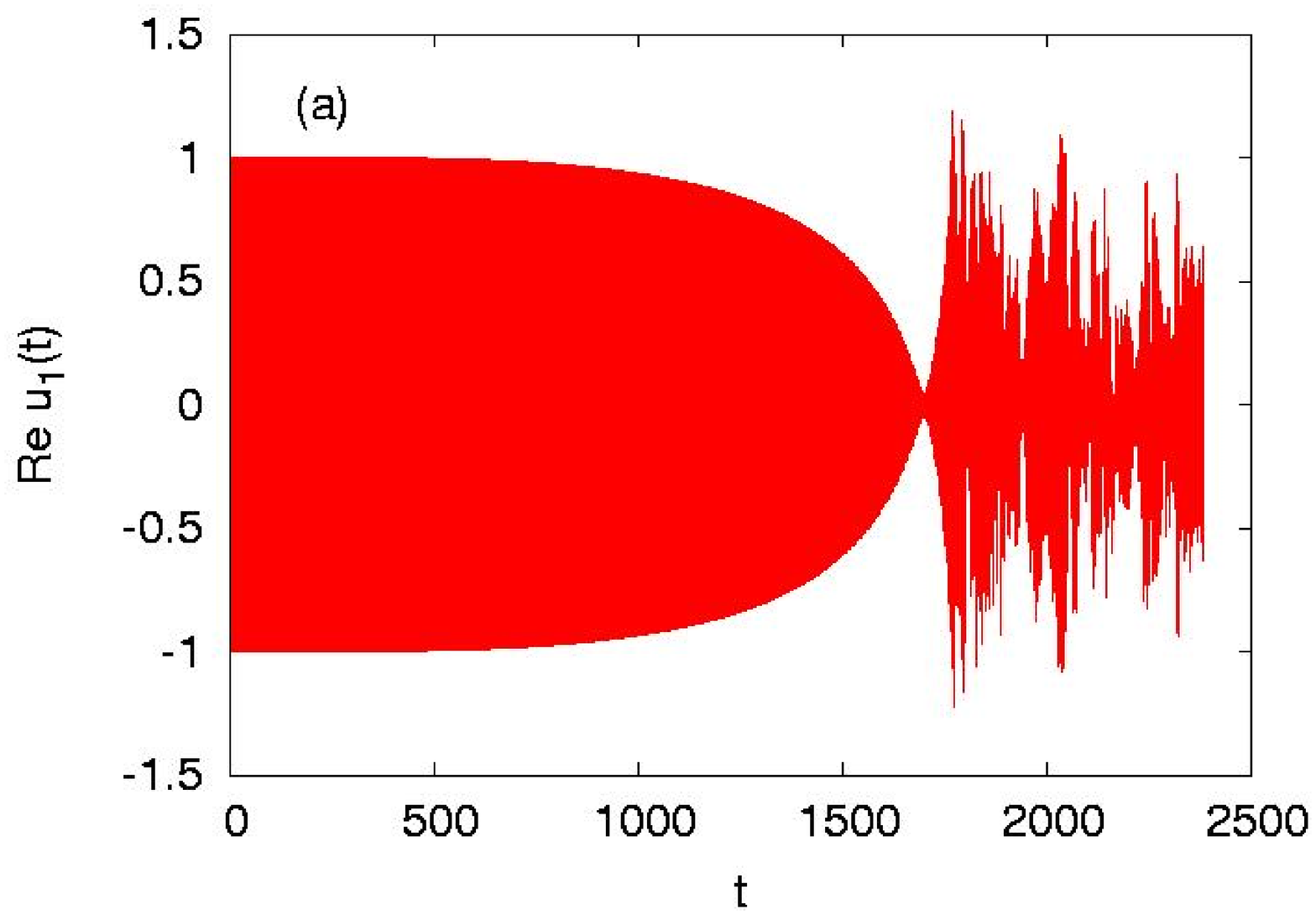,angle=-90}
}
{
\epsfig{width=.32\textwidth,file=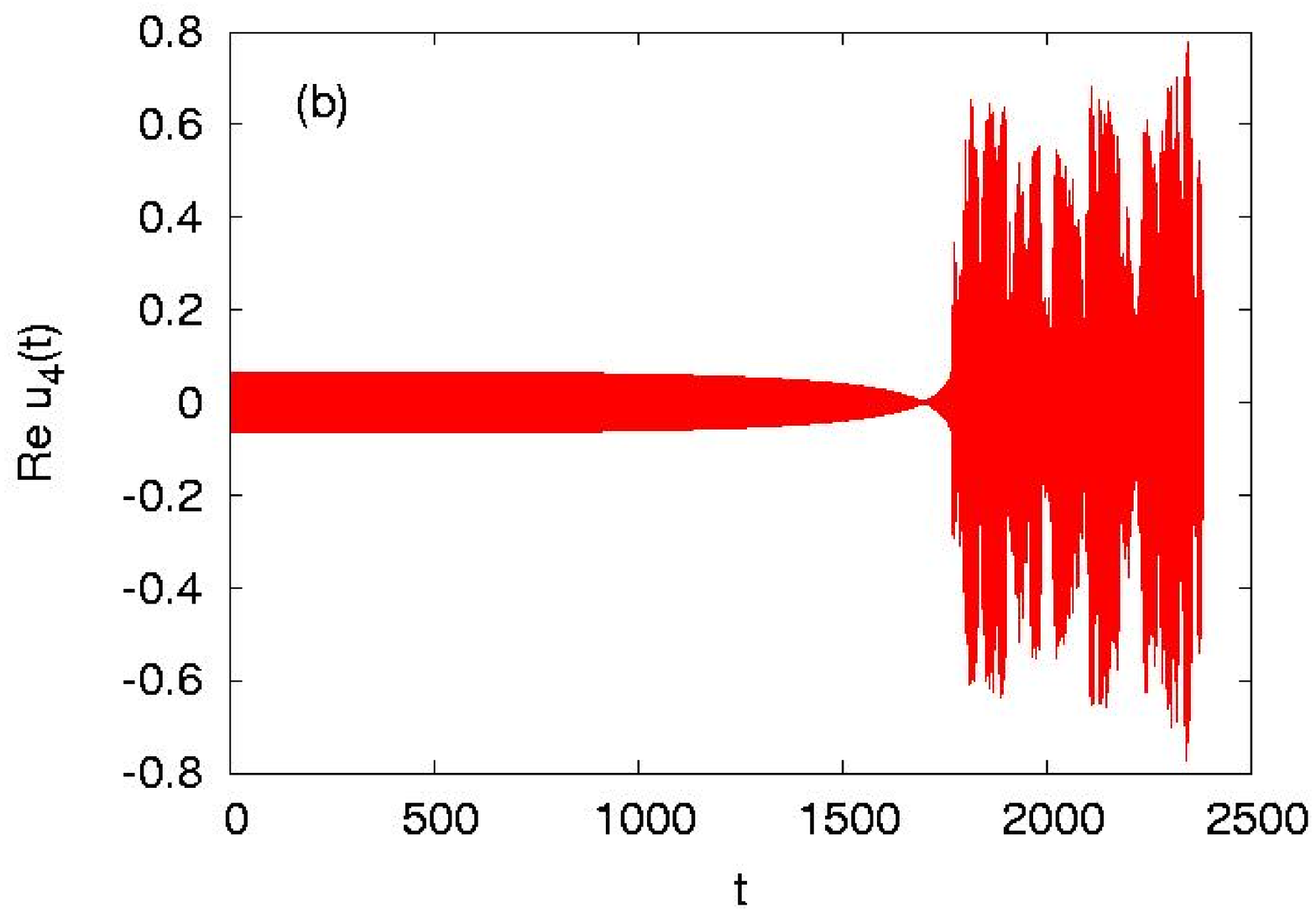,angle=-90}
}}
\caption{
Solutions to the GOY model with six shells and with the same
parameter values as in Fig. 4 but with addition of a small
forcing $\R {\it f} = \I {\it f} = 0.001$. 
Note that the solutions appear to be stable for a long time
but turn into chaotic trajectories eventually. 
Plotted are: a) $\R~ u_1(t)$; b): $\R~ u_4(t)$.
}
\label{fig:6}
\end{figure}

In order to study the stability of
the solutions further, we have initiated the integration
with the exact solutions shown in Fig. \ref{fig:4} but with the
addition of a small force. As is seen form the simulation
(shown in Fig. \ref{fig:6}),
the solution starts out to be stable, looking very similar
to the exact solution, however in the long run also become
unstable and chaotic. Nevertheless, we have observed numerically
that the existence of this transition depends very much on the 
value of the added forcing $f$. For some values of $f$, the
solutions remain stable although slightly changed away
from the exact solutions. For other values of $f$ we observe
the scenario depicted in Fig. \ref{fig:6}. We shall not
elaborate further on these transitions in the present paper. 

\begin{figure}[htbp]
\hbox{
{
\epsfig{width=.36\textwidth,file=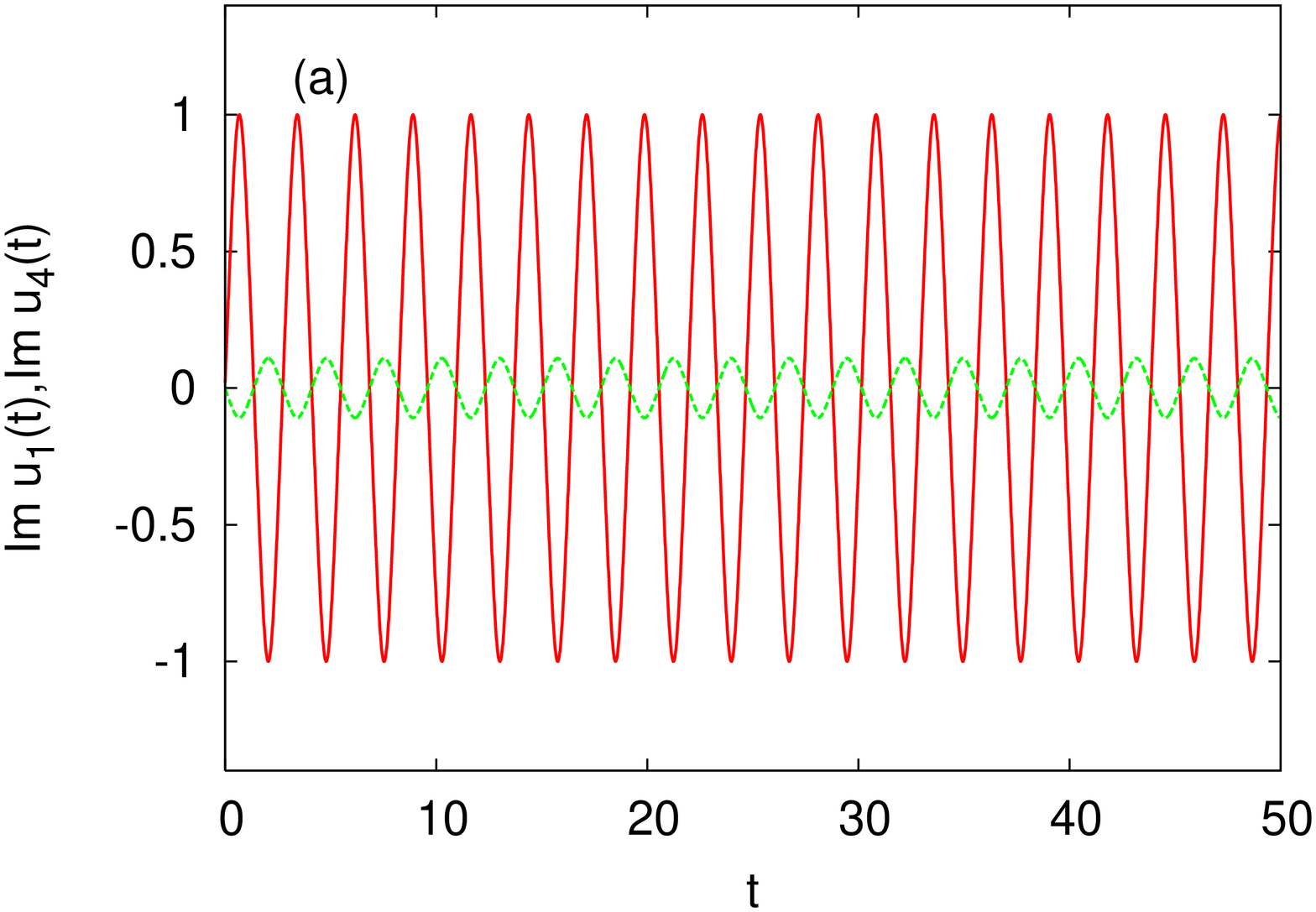,angle=-90}
}
{
\epsfig{width=.36\textwidth,file=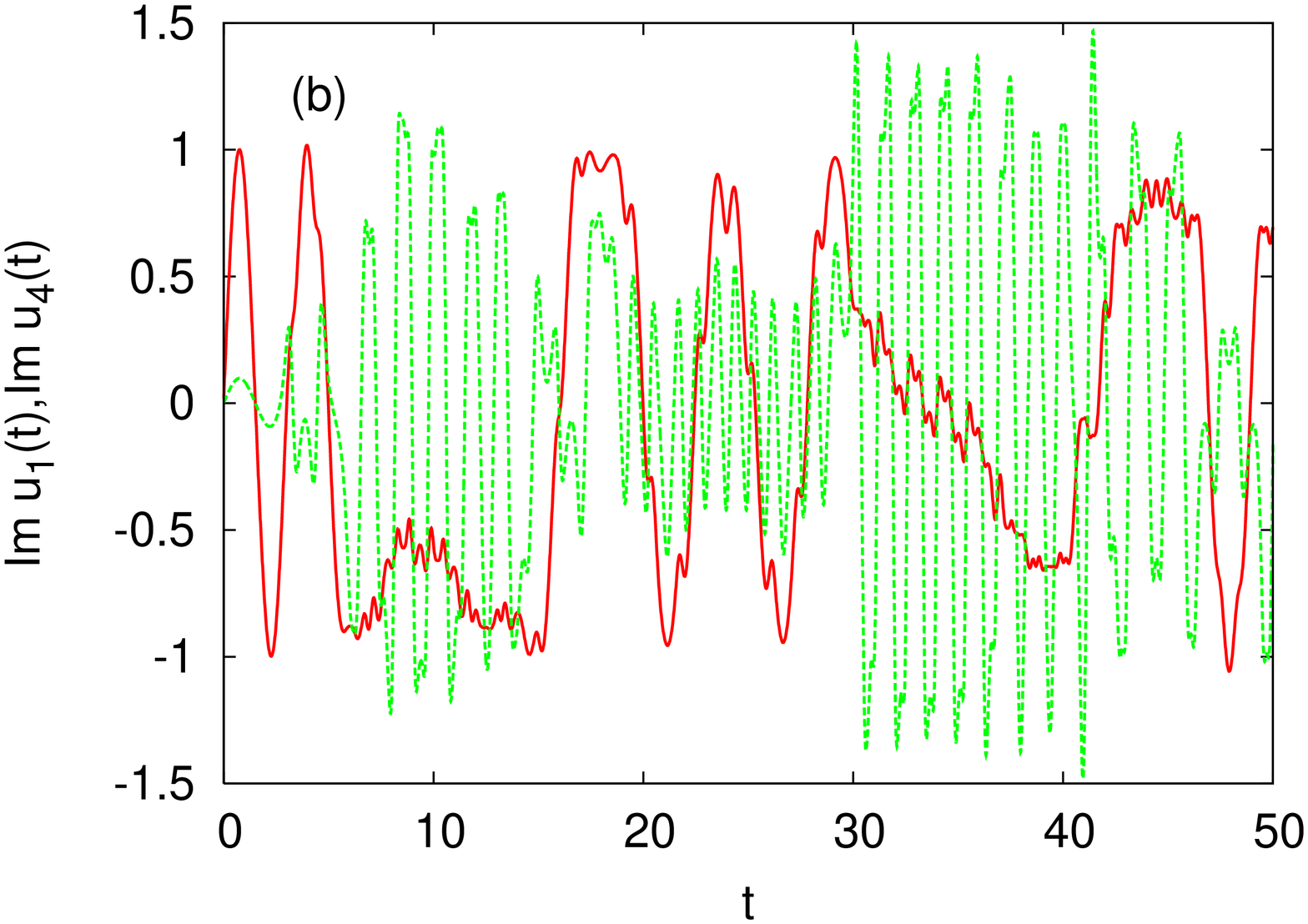,angle=-90}
}}
\caption{Plot of the Jacobi elliptic functions for six shells
for the case of $\delta=1.25$ for enstrophy conservation
corresponding to 2D turbulence for two different sets of
$\alpha_n$ values from (\ref{table2}) 
(the initial conditions are adjusted accordingly). a): $\alpha_4=-0.108771,~~~\alpha_5=0.112449,
~~~\alpha_6=-0.0517172$. Here the solutions in terms of elliptic
functions are stable.  b): $\alpha_4=0.0982932,~~~\alpha_5=0.359568,
~~~\alpha_6=0.485351$. The solutions 'follow' the elliptic solutions
for a few periods and the clearly become unstable.}
\label{fig:7}
\end{figure}

It is well known for the GOY shell model that at the
parameter value $\delta=1.25$ the enstrophy
$H = \sum_{n=1}^N k_n^2 |u_n|^2$ is a conserved quantity.
This point therefore corresponds to two-dimensional  turbulence
with a forward cascade of enstrophy and an inverse cascade 
of energy \cite{mogens}. We have therefore also looked for
exact solutions to the GOY model with six shells at this point and 
by using $r=2$ and $\delta=1.25$ we obtain as before from eqs. 
(\ref{c3}), (\ref{c4})
and (\ref{c5}) by use of NSolve in Mathematica the following 
real solutions:
\begin{eqnarray}
&&\alpha_4=0.289785,~~~\alpha_5=2.21803,
~~~\alpha_6=0.613153\nonumber \\
&&\alpha_4=-0.108771,~~~\alpha_5=0.112449,
~~~\alpha_6=-0.0517172,\nonumber \\
&&\alpha_4=0.0856125,~~~\alpha_5=0.160463,
~~~\alpha_6=0.302194,\nonumber \\
&&\alpha_4=0.0982932,~~~\alpha_5=0.359568,~~~\alpha_6=0.485351
\label{table2}
\end{eqnarray}
with the following relations between the amplitudes
$A_2$ and $A_1$
\begin{equation}
A_2^2/A_1^2= 0.67043, 1.46754, 1.07878, 1.05341
\end{equation}
and the elliptic parameter $k$ determined by:
\begin{equation}
k^2a^2/A_2^2=-4.9655, -0.473003, -0.0909193, -0.145639
\end{equation}
Of these four solutions to the 2D analogy for turbulence we
have found numerically that number one, three and four on the
list are unstable whereas number two is stable. As examples of
this we show in Fig. \ref{fig:7} some of the solutions, the ones corresponding to
the second and the fourth in the list. As compared to the 3D case, 
we observe that the solutions in Fig. \ref{fig:7}b become 
unstable even faster 
and that they clearly behave very chaotically. 

In Appendix A, we present solutions for the case of nine shells.
Furthermore, we derive a general recursion relation which allows
to solve the $\alpha_n$ coefficients for any number of shells.
If the number of shell goes to infinity one can see from these
recursion relations that the only scaling law solution
behaves like $\alpha_n \sim k_n^{-\frac{1}{3}}$ as expected from 
the Kolmogorov theory \cite{mogens}.

We end this section by pointing out that the solutions based on the
ansatz (\ref{ansatz5}) are special. Presumably the general recursion
relations (\ref{recursion}) 
for the amplitudes of the Jacobi functions can have 
more general 
solutions, where the constant phases could also enter the recursions.
In any case, the solutions that we found represent an infinite set of 
solutions, since the continuous parameters $A_1,~A_2,~a, ~b$ can vary freely.
Alternatively, an analysis of eqs. (\ref{ansatz}), (\ref{recursion}) shows
that the free parameters are $u_1(0), u_2(0), u_3(0)$ and $A_1$. 
Since these quantities in general are complex, this represents
an $\infty^8$-continuous set of parameters.

\section{Conclusions}

We have derived a series of exact analytical solutions to the GOY
shell model in the absence of forcing and viscosity. In this
model a shell couples to nearest and next-nearest neighbor
shells and the very construction of these couplings directly suggest 
that the solutions should be formulated in terms of Jacobi elliptic 
functions. We have presented closed form Jacobi solutions in the
case of three shells at which point the GOY model is completely
integrable and possesses a continuous infinity of stable solutions. 
We have further derived exact recursion relations for the amplitudes
of the elliptic functions in case of 6, 9, 12 etc shells which
is related to the period three symmetry of the GOY model. Some
of these solutions are stable and others are unstable. In the last
cases the trajectories lie on chaotic energy surfaces. This picture
of an infinity of solutions, some either periodic or quasi-periodic
and others chaotic is common for Hamiltonian systems where the
trajectories will move on surfaces of constant energy. Our results
therefore complement the picture presented for the GOY model with
forcing and viscosity obtained in \cite{GOYper} 
where long periodic unstable solutions 
are found embedded in the large dimensional phase space of the
dynamics of a GOY model with many shells. In this limit of many shells
we have found solutions that behave according to the Kolmogorov
scaling theory for energy spectra, which shows the physical relevance
of our periodic solutions. Indeed, we have also observed
that adding a small force, thus destroying the Hamiltonian property of
the system, changes the solutions. Some of them are destabilized but some
might remain periodic even when a force is added. The case of a forced
GOY model is closer to a realistic system for turbulence and it is
for future research to investigate how our solutions might be relevant 
for a turbulent state when a true energy cascade over many scales
is present.

\subsection*{Acknowledgments}
We are very grateful for discussions with Leo Kadanoff 
on shell models with few shells, which initiated this work.
Also we really appreciate the many very constructive comments by 
Bruno Eckhardt.

\section*{Appendix A: Solutions with nine shells}

In this appendix we shall discuss the case of nine shells. Following
the same procedure as in the main text
we obtain (the equations for $n<7$ are the same as before,
but for completeness we include the $n=6$ eq. (\ref{c3}) below):
\begin{equation}
\alpha_6=\frac{\alpha_4+r(1-\delta)}{r^3\alpha_5}+\frac{\delta}{r}.
\end{equation}
\begin{equation}
\alpha_7=-\frac{\alpha_5}{r^4\alpha_6}~(\delta-r\alpha_4)+\frac{\delta}{r}
\alpha_4+\frac{(1-\delta)~\alpha_4}{r^2\alpha_6},
\end{equation}
and
\begin{equation}
\alpha_8=-(-r^2\alpha_4\alpha_5+1-\delta+r\delta\alpha_4)\frac{\alpha_6}
{r^5\alpha_7}+\frac{\delta}{r}\alpha_5+\frac{1-\delta}{r^2\alpha_7}~
\alpha_5\alpha_4,
\end{equation}
and
\begin{equation}
\alpha_9=\frac{\alpha_7}{r^6\alpha_8}+\frac{\delta}{r}\alpha_6+
\frac{1-\delta}{r^2\alpha_8}\alpha_5\alpha_6,
\end{equation}
and
\begin{equation}
\frac{\delta}{r}~\alpha_7\alpha_9=\frac{\alpha_8}{r^7}~(\delta-r\alpha_4)
-\frac{1-\delta}{r^2}~\alpha_7\alpha_6,
\end{equation}
as well as
\begin{equation}
(1-\delta)\alpha_7\alpha_8=\frac{\alpha_9}{r^6}~(-r^2\alpha_4\alpha_5+1-\delta+
r\delta\alpha_4).
\end{equation}
These coupled equations can be solved using the simple Mathematica 
program ($\delta=.5, r=2$)
\begin{eqnarray}
&&{\rm NSolve}[\{z == (x + 1)/(2^3 y) + 1/4,
     r == -y (1/2 - 2 x)/(2^4 z) + 1/4 x + 1/2^3 x/z,\nonumber \\
  &&   s == -(-4 x y + 1/2 + x) z/(2^5 r) + 1/4 y + 1/2^3 x y/r,\nonumber \\
  &&t == r/(2^6 s) + 1/4 z + 1/2^3 y z/s,1/4 r t == s/2^7 (1/2 - 2 x) - 1/2^3 
z r,\nonumber \\
  &&   1/2 r s == t/2^6 (-4 x y + 1/2 + x)\}, \{x, y, z, r, s, t\}],
\label{program}
\end{eqnarray}
where the notation is
\begin{equation}
\alpha_4=x, ~\alpha_5=y,~\alpha_6=z,~\alpha_7=r,~\alpha_8=s,~\alpha_9=t,
\end{equation}
One obtains 32 solutions, some of which are complex, and some which leads
to $A_2^2<0$. Only seven of these solutions are acceptable. The reader
who is interested in the actual numbers should  just run the
above program. Increased accuracy can be obtained by putting $,d$ at the end
of the square bracket in (\ref{program}), where $d$ is the number of
wanted decimals.

For the $\alpha'$s we can also 
obtain in  general recursion relations 
by use of eq. (\ref{recursion}). Defining
$A_n=\alpha_n A_1$ if $n=4$ mod(3),$A_n=\alpha_n A_2$ if $n=5$ mod(3) and
$A_n=\alpha_n A_3$ if $n=6$ mod(3), we obtain for $n\geq 4$
\begin{equation}
\alpha_{n+2}=\frac{\tilde{\epsilon}_n\alpha_n}{r^{n-1}\alpha_{n+1}}+\frac
{\delta}{r}~\alpha_{n-1}+\frac{1-\delta}{r^2}~\frac{\alpha_{n-1}\alpha_{n-2}}
{\alpha_{n+1}},
\label{KA}
\end{equation}
where we defined $\alpha_1=\alpha_2=\alpha_3=1$. Here ($n=0,1,2,3,...$)
\begin{equation}
\tilde{\epsilon}_{4+3n}=1,~\tilde{\epsilon}_{5+3n}=-(\delta-r\alpha_4),~ 
\tilde{\epsilon}_{6+3n}=-(1-\delta+r\delta\alpha_4-r^2\alpha_4\alpha_5).
\end{equation}
In addition there are two ``stop-equations'': Suppose we stop at
the number of shells $=3N$, then we must satisfy that $A_{3N+1}=0$
and $A_{3N+1}A_{3N+2}=0$, which lead to the two equations
\begin{equation}
\delta~\alpha_{3N-2}\alpha_{3N}=\frac{\alpha_{3N-1}}{r^{3N-3}}~(\delta-
r\alpha_4)-\frac{1-\delta}{r}~\alpha_{3N-3}\alpha_{3N-2},
\label{KB}
\end{equation}
and
\begin{equation}
(1-\delta)~\alpha_{3N-2}\alpha_{3N-1}=\frac{\alpha_{3N}}{r^{3N-3}}~
(1-\delta+r\delta\alpha_4 -r^2\alpha_4\alpha_5).
\label{KC}
\end{equation}
By means of eqs. (\ref{KA}), (\ref{KB}) and (\ref{KC}) it is
easy to write down the relevant equations for any number of shells.
The remaining problem is the numerical solution of these equations. We
are not able to show the existence of acceptable solutions in general, 
but have checked that they exist for six and nine shells.

In case with an infinite number of shells, $N \to \infty$, 
the ``stop-equations'' (\ref{KB}) and (\ref{KC}) should
of course not be applied so we are left only with eq. (\ref{KA}).
If we make the ansatz $\alpha_n \sim r^{\beta_n}$, then the only
solution is
\begin{equation}
\beta_n = - \frac{n}{3}~~,
\end{equation}
which is fully consistent with the Kolmogorov theory for
turbulence. The first term in Eq. (\ref{KA}) is totally
subdominant with this ansatz.

\section*{Appendix B: Solutions to the Sabra model}

In this appendix we shall derive results similar to those in Section III
for the Sabra-model \cite{sabra1}. This model is given by
\begin{equation}
\dot{u}_n=ir^n \left[u_{n+1}^\star u_{n+2}-\frac{\delta}{r}u_{n-1}^
\star u_{n+1}+\frac{1-\delta}{r^2}u_{n-1}u_{n-2}\right].
\end{equation}
We can now easily check that if the ansatz (\ref{ansatz}) is inserted
in this equation, like in the GOY model, the Jacobi functions cancel on 
both sides and we
are left with the following recursion relation
\begin{equation}
A_{n+2}=-\frac{ia\epsilon_nA_n}{r^{n-1}A^\star_{n+1}}+\frac{\delta}{r}~A^
\star_{n-1}~\frac{A_{n+1}}{A^\star_{n+1}}-\frac{1-\delta}{r^2}~\frac
{A_{n-1}A_{n-2}}{A_{n+1}^\star}.
\label{bbb100}
\end{equation}
This is the Sabra-version of the analogous GOY model equation 
(\ref{recursion}). We can
then proceed to solve for the first few $A'$s,
\begin{equation}
A_3=-\frac{iaA_1}{A_2^\star},
\end{equation}
which is similar to eq. (\ref{aaa3}),
\begin{equation}
A_4=-\frac{A_2}{rA_1^\star A_2^\star}~\left(\delta |A_1|^2-|A_2|^2\right),
\end{equation}
similar to eq. (\ref{a4}), and
\begin{equation}
A_5=\frac{A_1^2A_2}{r(A_2^\star)^2|A_1|^2(\delta |A_1|^2-|A_2|^2)}
\left[-a^2k^2|A_1|^2+(\delta^2-\delta+1)|A_1|^2|A_2|^2+\delta |A_2|^4\right],
\end{equation}
which is similar to eq. (\ref{a5}). One can of course continue
like in Section III to obtain further equations from the general recursion
relation (\ref{bbb100}). 

For the case of three shells we should again require that $A_4=0$, which
gives
\begin{equation}
|A_2|^2=\delta |A_1|^2.
\label{bbb101}
\end{equation}
Requiring further that $A_5A_4=0$, we obtain
\begin{equation}
a^2k^2=(1-\delta)~|A_2|^2.
\label{bbb102}
\end{equation}
Eqs. (\ref{bbb101}) and (\ref{bbb102}) are completely the same as 
the analogous equations obtained in the GOY model in Section III.
However, in the case of more shells this  is certainly not
expected to continue.

\end{document}